\documentclass[aps,prd,onecolumn,nofootinbib,showkeys]{revtex4-1} 

\usepackage[T1]{fontenc}
\usepackage[utf8]{inputenc}
\usepackage{amsmath,amssymb,amsfonts}
\usepackage{graphicx,epsfig}
\usepackage{color}
\usepackage{hyperref}
\usepackage{array}
\usepackage{subfigure}  
\usepackage{multirow}
\usepackage{float}
\usepackage{booktabs}

\begin{document}

\title{Asymmetric thin-shell wormholes in the Kalb-Ramond background: Observational characteristics and extra photon rings}

\author{K.TAN}

\author{X.G.Lan}
\email{E-mail: xglan@cwnu.edu.cn}
\affiliation{ Institute of Theoretical Physics, China West Normal University, Nanchong 637009, China}

\date{\today}

\begin{abstract}

In this paper, we utilize the ray-tracing method to conduct an in-depth study of the observational images of asymmetric thin-shell wormholes in the Kalb-Ramond field. Initially, we calculate the null geodesics and effective potential energy of the asymmetric thin-shell wormhole, and investigate the variations in the photon sphere radius and critical impact parameter under different values of charge $Q$ and Lorentz-violation parameter $l$. Based on these calculations, we determine the photon deflection angles and trajectories within this space-time structure. Specifically, depending on the photon impact parameters, the photon trajectories can be categorized into three types. By using a thin accretion disk as the sole background light source and incorporating two classical observational radiation models, we find that under conditions of equal mass $M$, charge quantity $Q$, and Lorentz-violation parameter $l$, the asymmetric thin-shell wormholes exhibit unique observational features such as additional lensing rings and photon ring clusters. Furthermore, distinct from black holes, as the charge quantity $Q$ and the Lorentz-violation parameter $l$ increase, the coverage area of the specific additional halo also expands correspondingly.

\end{abstract}

\keywords{Kalb-Ramond, asymmetric thin-shell wormhole,  observational characteristics, Lorentz-violation}

\maketitle
\tableofcontents  

\section{Introduction}
In 2015, the Laser Interferometer Gravitational-Wave Observatory (LIGO) made the first detection of gravitational waves, directly confirming the existence of binary stellar-mass black holes-a landmark achievement in astrophysics~\cite{LIGOScientific:2016sjg}. Subsequently, in 2019, the Event Horizon Telescope (EHT) released images of the supermassive black holes at the centers of the M87 galaxy and Sgr A~\cite{EventHorizonTelescope:2019dse}, providing further evidence for the existence of such supermassive black holes. The black hole images display a central dark region surrounded by a bright ring: the dark area results from the black hole's intense gravity bending and capturing light, preventing it from reaching distant observers, thus forming what is known as the black hole shadow. Within the shadow, the bright ring contains a very thin, embedded ring—commonly referred to as the photon ring—which theoretically corresponds to the so-called "critical curve"~\cite{Gralla:2019xty}.

The study of black hole shadows has always been a hot topic in the field of physics~\cite{Perlick:2021aok,Zhang:2024lsf,Zeng:2025olq,Hou:2022gge,Wang:2025ihg,Zhang:2025vyx,He:2024amh,Fan:2024rsa,Zeng:2025kqw}. Synge was the first to calculate and determine the boundary of the black hole shadow, while Luminet, pioneered the study of the angular radius equation for the photon capture region in a Schwarzschild black hole~\cite{Synge:1966okc,Luminet:1979nyg}. Falcke demonstrated that the shadow of Sgr A* could be observed via very long-baseline interferometry at sub-millimeter wavelengths, under the assumption that the accretion flow is optically thin in this spectral region~\cite{Falcke:1999pj}. Additionally, extensive research has also been conducted on the signature of  modified theory of gravity on the photon sphere and shadow, and some meaningful progress has been achieved~\cite{Guo:2018kis,Wang:2018prk}.
 Although the EHT's observations are largely consistent with the black hole shadow predictions of General Relativity, they do not rule out the existence of certain Ultra-Compact Objects (UCOs)~\cite{Bambi:2008jg,Banerjee:2019nnj,Abdikamalov:2019ztb}. In fact, objects such as wormholes and boson stars could potentially cast shadows very similar to those of black holes~\cite{Tsukamoto:2021fpp,Wielgus:2020uqz,Wang:2023nwd,Rosa:2023qcv,Rosa:2022tfv,Cunha:2016bjh} . Consequently, finding methods to effectively distinguish UCOs from black holes has become a crucial research direction.

 Interestingly, Matt Visser proposed a novel construction known as the Thin-Shell Wormhole, which connects two distinct space-times via a throat, allowing for potential traversal~\cite{Visser:1989kh,Visser:1989kg}. Building upon this theoretical framework, Wang  investigates the optical appearance of an asymmetric thin-shell wormhole(ATSW)~\cite{Wang:2020emr}, and reveale that its shadow is consistently smaller than that produced by a black hole-a significant result that provides favorable support for the direct observation of wormholes. Subsequent research has primarily focused on the double shadow phenomenon generated by ATSWs. The presence of this double shadow may offer an alternative method for indirectly observing ATSW~\cite{Guo:2022iiy,Chen:2022tog}. Observational results of these dual shadows could not only test predictions of General Relativity but also enhance our understanding of gravitational effects and space-time curvature~\cite{Guerrero:2021pxt,Wang:2024uda,Guo:2022nto,Huang:2024wpj,Wei:2023fkn}. Furthermore, Peng et al. concentrated on the observational signatures of accretion disks and the additional photon rings around a Schwarzschild ATSW~\cite{Peng:2021osd}. These features can likewise serve as a method to distinguish ATSWs from black holes. Beyond this, other researchers have studied the observational appearance of wormholes in Hayward space-time and the appearance of a freely falling star within an ATSW~\cite{Jing:2024thh,Chen:2024wtw,He:2024yeg,Luo:2023wru}. Inspired by these studies, this paper will investigate the observational characteristics of an ATSW surrounded by an accretion disk within the context of Kalb-Ramond(KR) gravity.

The structure of this paper is as follows. In Section~\ref{sec2}, an ATSW is constructed within the KR field. We investigate its geodesics and effective potential, and subsequently calculate the deflection and trajectories of photons in this ATSW space-time. Section~\ref{sec3} examines the transfer functions and observational appearances of both a BH and an ATSW, each surrounded by a thin accretion disk and under the same mass parameter, for two emission models. Finally, our conclusions are presented in Section~\ref{sec4}.

\section{Null Geodesics and Effective Potential of the ATSW in the Kalb-Ramond Field}\label{sec2}
In the 1970s, Kalb and Ramond introduced the concept of an antisymmetric tensor field~\cite{Kalb:1974yc}. The KR field is a second-rank antisymmetric tensor~\cite{Daniali:2025yfz,Capanelli:2023uwv,Jumaniyozov:2025dyy,Zahid:2024hyy,Shodikulov:2025xax,Yang:2025byw}. It plays a significant role in theoretical physics by not only extending the understanding of gauge fields in quantum field theory but also being applied to advanced research areas such as string theory and dark matter. In recent years, substantial progress has been made in this field, with researchers having derived Schwarzschild-like solutions and rotating black hole solutions within the KR field framework~\cite{Duan:2023gng}. Our study will commence from the action of the KR field to conduct further investigation
\begin{equation}\label{eq1}
\begin{aligned}
S=  \frac{1}{2} \int \mathrm{~d}^4 x \sqrt{-g} \left[ R - 2 \Lambda - \frac{1}{6} H^{\mu v \rho} H_{\mu v \rho} \right. \left. - V\left(B^{\mu v} B_{\mu v} \pm b^2\right) \right] + \xi_1 B^{\rho \mu} B_\mu^v R_{\rho v} + \xi_2 B^{\mu v} B_{\mu v} R + \int \mathrm{d}^4 x \sqrt{-g} \mathcal{L}_{\mathrm{N}},
\end{aligned}
\end{equation}

where
\begin{equation}\label{eq2}
\begin{gathered}
B^{\mu \nu} B_{\mu \nu}=\mp b^2, \\
\mathcal{L}_{\mathrm{N}}=-\frac{1}{2} F^{\mu \nu} F_{\mu \nu}-\eta B^{\alpha \beta} B^{\gamma \rho} F_{\alpha \beta} F_{\gamma \rho}, \\
F_{\mu \nu}=\partial_\mu A_\nu-\partial_\nu A_\mu,\\
\tilde{H}^{\mu \nu \rho} \tilde{H}_{\mu \nu \rho}=H^{\mu \nu \rho} H_{\mu \nu \rho}+2 H^{\mu \nu \rho} A_{[\mu} F_{\nu \rho]}+A^{[\mu} F^{\nu \rho]} A_{[\mu} F_{\nu \rho]}.
\end{gathered}
\end{equation}

In the theoretical framework, $R$ denotes the Ricci scalar, $\eta$ and $\xi_1$, $\xi_2$  are coupling constants representing the non-minimal coupling terms between gravity and the KR field, while $\mathcal{L}_{\mathrm{N}}$ refers to the Lagrangian of the electromagnetic field. In our previous work, we investigated the influence of the charge $Q$ and the Lorentz-violation parameter $l$ on black hole shadows~\cite{Tan:2025pya}. Specifically, the study focused on a KR black hole surrounded by a thin accretion disk. It was found that an increase in the charge $Q$ and the Lorentz-violation parameter $l$ leads to a reduction in the size of the black hole shadow, as well as in the direct image, lensing ring, and photon ring. The present research shifts focus to the optical characteristics of an Asymmetric Thin-Shell Wormhole within the KR framework. This investigation is motivated by two key aspects: First, previous studies on Bardeen-type ATSWs have shown that the magnetic charge $g$ significantly affects their optical appearance~\cite{He:2024yeg}. This intriguing result prompts us to examine whether the charge $Q$ and the Lorentz-violation parameter $l$ similarly induce changes in the optical appearance of the KR field ATSW, analogous to the case of the KR black hole~\cite{Zeng:2024ptv}. Second, we aim to determine whether shadow-related characteristics can provide a viable method for distinguishing an ATSW from a black hole.

The ATSW consists of two space-times with distinct mass parameters, $M_1$ and $M_2$, connected by a thin shell (ie, the throat), with the entire ATSW manifold denoted by $M=M_1 \cup M_2$. Based on Eq.~(\ref{eq1}), with the cosmological constant $\Lambda$ set to zero, we can derive the following equation~\cite{Duan:2023gng}
\begin{equation}\label{eq3}
\begin{aligned}
\mathrm{d} s_i^2 = & -F_i\left(r_i\right) \mathrm{d} t_i^2 + \frac{1}{F_i\left(r_i\right)} \mathrm{d} r_i^2 + r_i^2\left(\mathrm{~d} \theta_i^2 + \sin^2 \theta_i \mathrm{~d} \phi_i\right).
\end{aligned}
\end{equation}
The metric function can be expressed as
\begin{equation}\label{eq4}
F_i\left(r_i\right)=\frac{1}{1-l}-\frac{2 M_i}{ r_i}+\frac{Q^2}{(1-l)^2  r_i^2}.
\end{equation}
Here, $M$, $Q$, and $l$ represent the black hole's mass, charge, and Lorentz-violation parameter, respectively. Expanding Eq.~(\ref{eq4}) as a Taylor series in small $l$ yields
\begin{equation}\label{eq5}
\begin{aligned}
F_i\left(r_i\right) = & \left(1 + \frac{Q^2}{r_i^2} - \frac{2 M_i}{r_i}\right) + \left(1 + \frac{2 Q^2}{r_i^2}\right) l  + \left(1 + \frac{3 Q^2}{r_i^2}\right) l^2, \quad r_i \geq R.
\end{aligned}
\end{equation}
In which, $i=1, 2$ denotes the two space-time regions, $R$ represents the position of the thin shell and satisfies the condition $R>\max \left\{r_{h_1}, r_{h_2}\right\}$, where $r_{h_1}$ and $r_{h_2}$ represent the event horizon radii of black holes with mass parameters $M_1$ and $M_2$, respectively. Considering that gravity is the sole interaction as photons traverse the throat, the energy $p_t$ and axial angular momentum $p_\phi$ are all conserved.

Since the space-time is symmetric, we choose the plane $\theta 
= \pi/2$ (passing through the coordinate origin) for analysis; photons then possess two conserved quantities: $E_i=-p_{t_i}$
(energy) and $L_i=p_{\phi_i}$ (angular momentum) when propagating along geodesics, and their motion satisfies the geodesic equations
\begin{equation}\label{eq6}
\frac{\left(p_i^{r_i}\right)^2}{F_i\left(r_i\right)}+\frac{p_{\phi_i}^2}{r_i^2}=\frac{p_{t_i}^2}{F_i\left(r_i\right)},
\end{equation}
where $p_i^{r_i}=\mathrm{d} x^{r_i} / \mathrm{d} \lambda$ represents the four-momentum of the photon in the space-time, and $\lambda$ represents the affine parameter. 
It is conventional to redefine the affine parameter using the conserved energy as $\lambda=\lambda / E$, which allows the impact parameter $b$ — a dimensionless quantity characterizing the photon's trajectory — to be precisely defined as the ratio of the conserved angular momentum to the conserved energy: $b=|L| / E$~\cite{Luo:2023wru,He:2024yeg}. Based on these relations, we derive the following expression
\begin{equation}\label{eq7}
p_i^{r_i}= \pm E_i \sqrt{1-\frac{b_i^2}{r_i^2} F_i\left(r_i\right)}.
\end{equation}
In Eq.~(\ref{eq7}), $+$ and $-$ represent the photon's outgoing and incoming directions, respectively. Consequently, we can derive the effective potential for an ATSW in the KR field
\begin{equation}\label{eq8}
\begin{aligned}
V_{e\!f\!f_i}\left(r_i\right) = & \frac{F_i\left(r_i\right)}{r_i^2} = \frac{1}{r_i^2}\left(1 + \frac{Q^2}{r_i^2} - \frac{2 M_i}{r_i}\right) + \frac{1}{r_i^2}\left(1 + \frac{2 Q^2}{r_i^2}\right) l + \frac{1}{r_i^2}\left(1 + \frac{3 Q^2}{r_i^2}\right) l^2.
\end{aligned}
\end{equation}
For a photon to move on a circular orbit, the orbit must satisfy the conditions for an extremum of the effective potential
\begin{equation}\label{eq9}
V_{e\!f\!f_i}\left(r_{p h_i}\right)=\frac{1}{b_{c_i}^2}, \quad V_{e\!f\!f_i}^{\prime}\left(r_{p h_i}\right)=0,\quad
\frac{d^2 V_{e\!f\!f_i}(r_{p h_i})}{dr^2} \Big|_{r = r_{p h_i}} < 0.
\end{equation}

Here, $b_{c_i}$ and $r_{p h_i}$ denote the critical impact parameter and the photon sphere radius, respectively. When the condition \(b = b_c\) is satisfied — meaning the light ray’s impact parameter is infinitely close to the radius of the photon sphere — the ray will orbit the black hole infinitely many times.

The stability of this orbit is determined by the sign of the second derivative. An unstable circular orbit, which defines the photon sphere, corresponds to a local maximum. By combining Eqs.~(\ref{eq5}) and ~(\ref{eq9}), we obtain the expression for $r_{p h_i}$ and $b_{c_i}$
\begin{equation}
r_{p h_i}=\frac{3 M_i+\sqrt{9 M_i^2-4\mathcal{AA}\left(2 Q^2+4 Q^2 l+6 Q^2 l^2\right)}}{2\mathcal{AA}},
\end{equation}
\begin{equation}
b_{c_i}=\frac{\sqrt{-27 M_i^4+36 M_i^2 Q^2 \mathcal{BB}-8 Q^4 \mathcal{BB}^2-9 M_i^3 \sqrt{9 M_i^2-8 O^2 \mathcal{BB}}+8 M_i Q^2 \mathcal{BB} \sqrt{9 M_i^2-8 Q^2 \mathcal{BB}}}}{\sqrt{2} \sqrt{\mathcal{AA}\left(-M_i^2+Q^2\mathcal{BB}\right)}},
\end{equation}
where
\begin{equation}
\mathcal{AA}=1+l+l^2,  \mathcal{BB}=1+3 l+6 l^2+5 l^3+3 l^4.
\end{equation}

This leads to the expression for the area $S_{p_i}$ enclosed by the photon sphere radius in the two-dimensional plane
\begin{equation}
S_{p_i} = \pi r_{p h_i}^2.
\end{equation}
The primary objective of this paper is to investigate a method for distinguishing an ATSW from a black hole within the KR field, specifically under the condition where the event horizon is enclosed by the photon sphere. To this end, we assume an observer is situated in the space-time with mass parameter $M_1=1$, and the mass parameter of the other space-time $M_2$ is set to $M_2=k$. Under this configuration, the parameters $k$ and the throat radius $R$ must satisfy the following conditions~\cite{Wang:2020emr,Peng:2021osd}
\begin{equation}\label{eq10}
1<k<\frac{R}{2} \leq \frac{r_{p h_1}}{2}.
\end{equation}
In essence, within the valid range of the parameter $k$, its specific value does not qualitatively affect the subsequent results and discussion. Therefore, we adopt the approach used in prior studies and set the value of 
$k=1.2$~\cite{Peng:2021osd,Guo:2022iiy}. 
The local tetrad frame in the neighborhood of the thin-shell wormhole in spacetime $M_{1,2}$ can be expressed as
\begin{equation}\label{eqA}
\mathbf{e}_{t_{i}}^{a} \equiv F_{i}^{-\frac{1}{2}}(R)\left(\frac{\partial}{\partial t_{i}}\right)^{a}, \quad 
\mathbf{e}_{r_{i}}^{a} \equiv \sqrt{F_{i}}(R)\left(\frac{\partial}{\partial r_{i}}\right)^{a},
\end{equation}
which are then related by the Lorentz transformation~\cite{Langlois:2001uq}
\begin{equation}
    \begin{pmatrix}
        \mathbf{e}_{t_{2}}^{a} \\
        \mathbf{e}_{r_{2}}^{a}
    \end{pmatrix}
    = \Lambda(\theta_1 - \theta_2)
    \begin{pmatrix}
        \mathbf{e}_{t_{1}}^{a} \\
        \mathbf{e}_{r_{1}}^{a}
    \end{pmatrix}.
\end{equation}
We define $\Lambda(\theta_i)$ and $\theta_i$ within the range as follows
\begin{equation}
    \begin{aligned}
        \Lambda(\theta_i) &= \begin{pmatrix}
            \cosh\theta_i & \sinh\theta_i \\
            \sinh\theta_i & \cosh\theta_i
        \end{pmatrix}, \quad
        \theta_i = \sinh^{-1} \left( \frac{\epsilon_i \dot{R}}{\sqrt{F_i(R)}} \right).
    \end{aligned}
\end{equation}

\begin{figure*}[htbp]
	\centering
	\subfigure[]{
		\begin{minipage}[t]{0.28\linewidth}
			\includegraphics[width=2.1in]{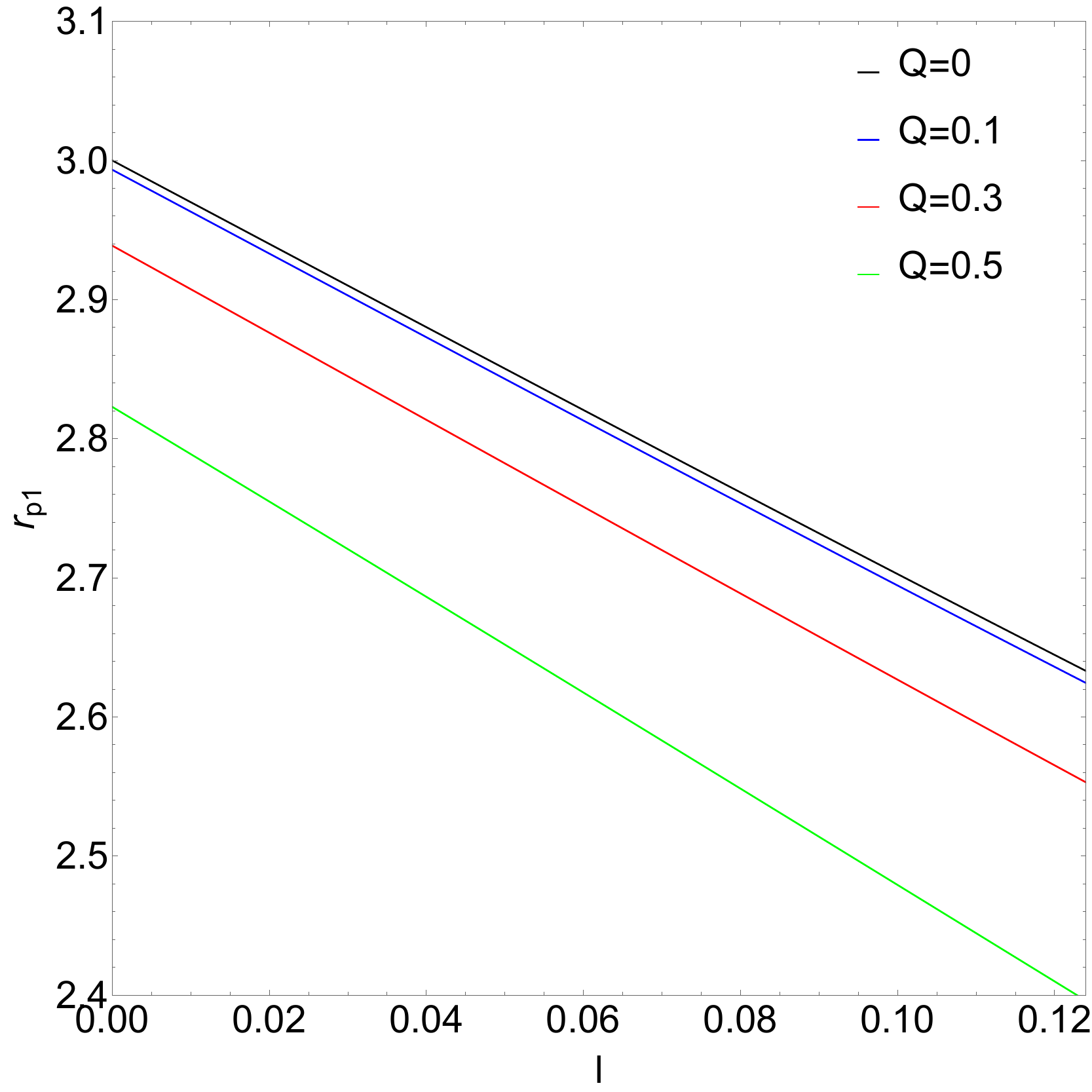}\\
		\end{minipage}
	}
	\subfigure[]{
		\begin{minipage}[t]{0.28\linewidth}
			\includegraphics[width=2.1in]{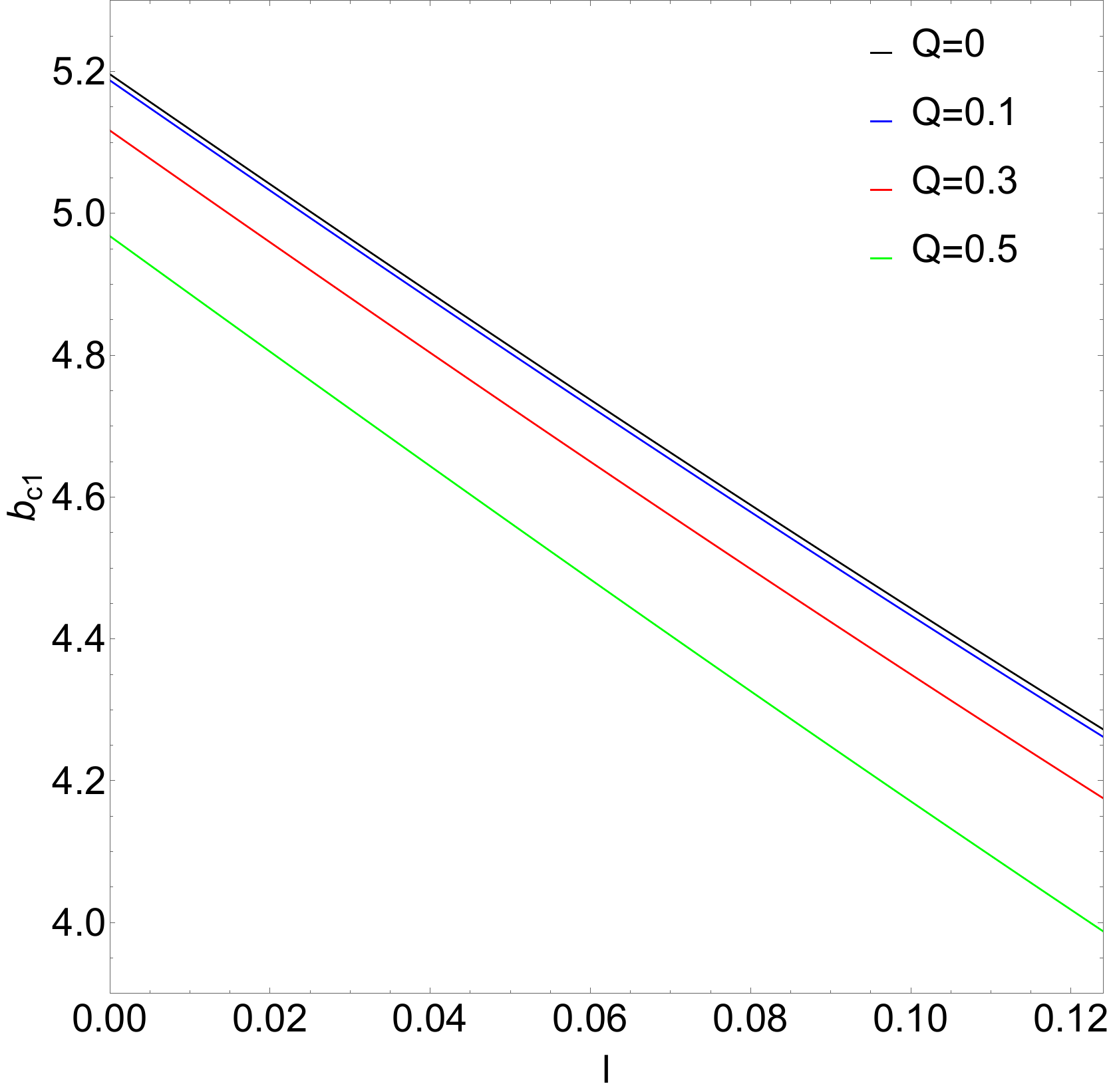}\\
		\end{minipage}
	}
    \subfigure[]{
		\begin{minipage}[t]{0.28\linewidth}
			\includegraphics[width=2.1in]{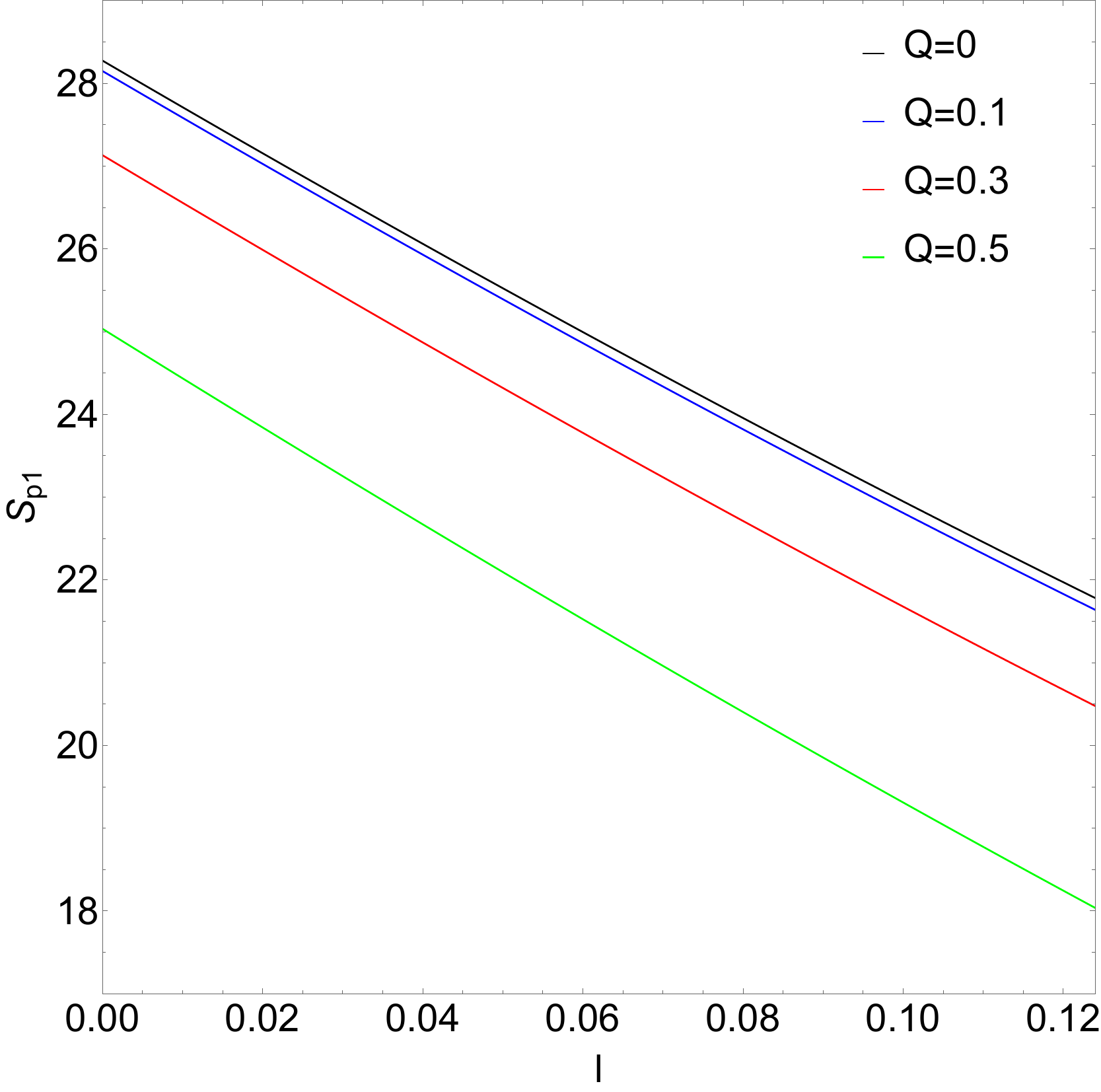}\\
		\end{minipage}
	}
	\caption{(color online) In spacetime \(M_1\), panels (a), (b), and (c) respectively depict the variation of the photon sphere radius \(r_{p_{1}}\), impact parameter \(b_{c_{1}}\), and the enclosed area \(S_{p_{1}}\) of the photon sphere with the Lorentz-violation parameter $l$.}
	\label{S1}
\end{figure*}

\begin{figure*}[htbp]
	\centering
	\subfigure[]{
		\begin{minipage}[t]{0.28\linewidth}
			\includegraphics[width=2.1in]{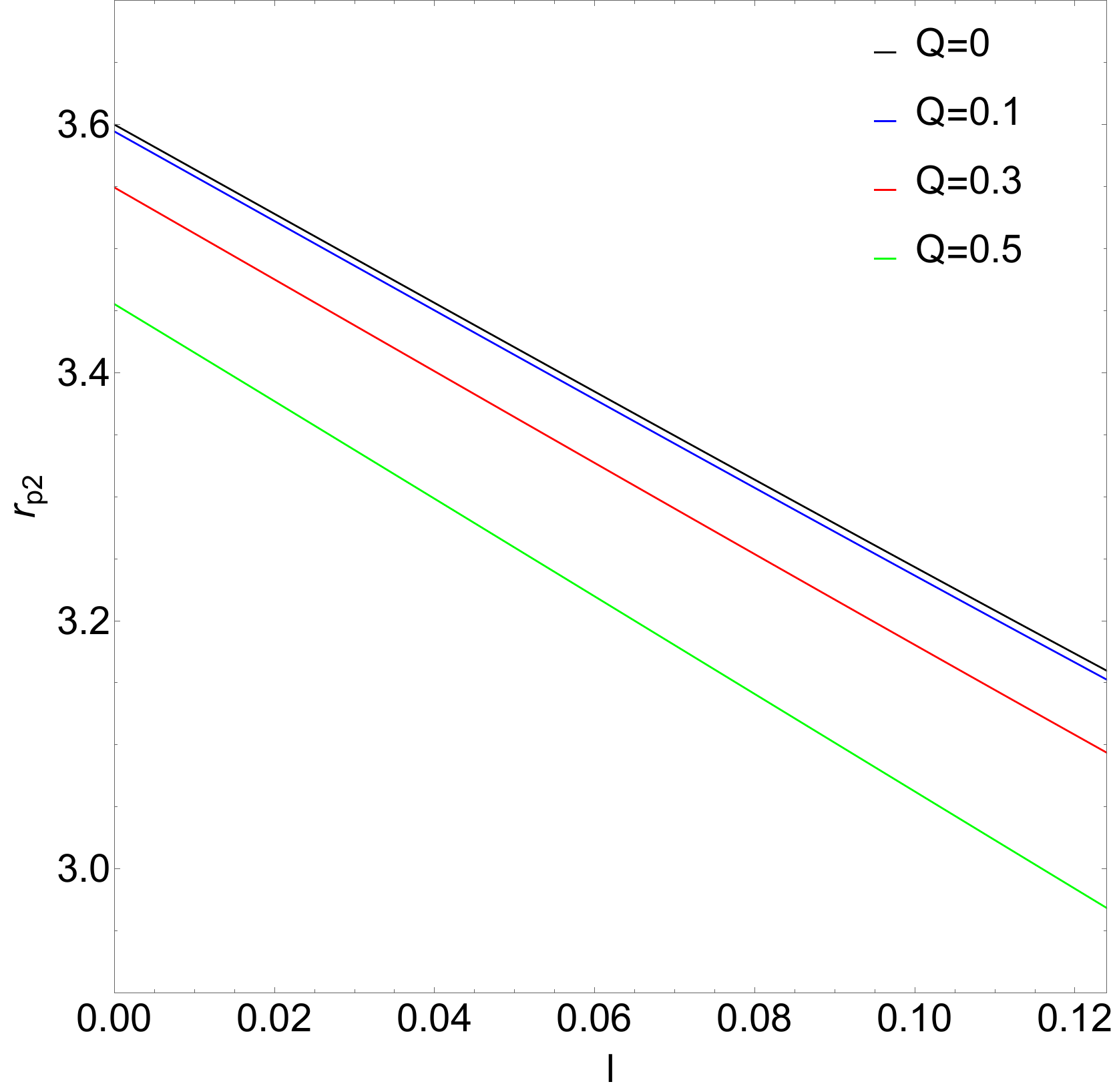}\\
		\end{minipage}
	}
	\subfigure[]{
		\begin{minipage}[t]{0.28\linewidth}
			\includegraphics[width=2.1in]{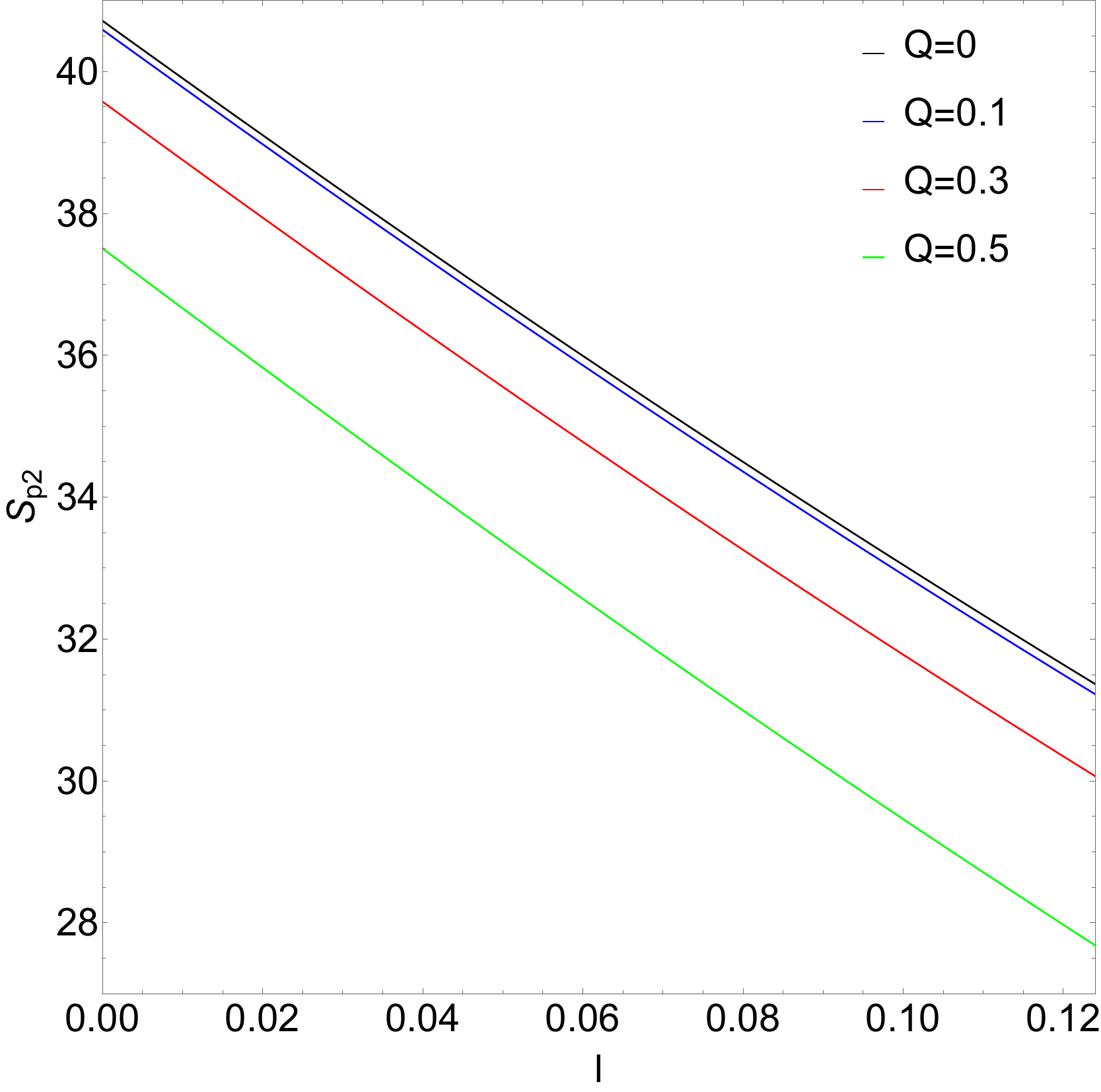}\\
		\end{minipage}
	}
	\subfigure[]{
		\begin{minipage}[t]{0.28\linewidth}
			\includegraphics[width=2.1in]{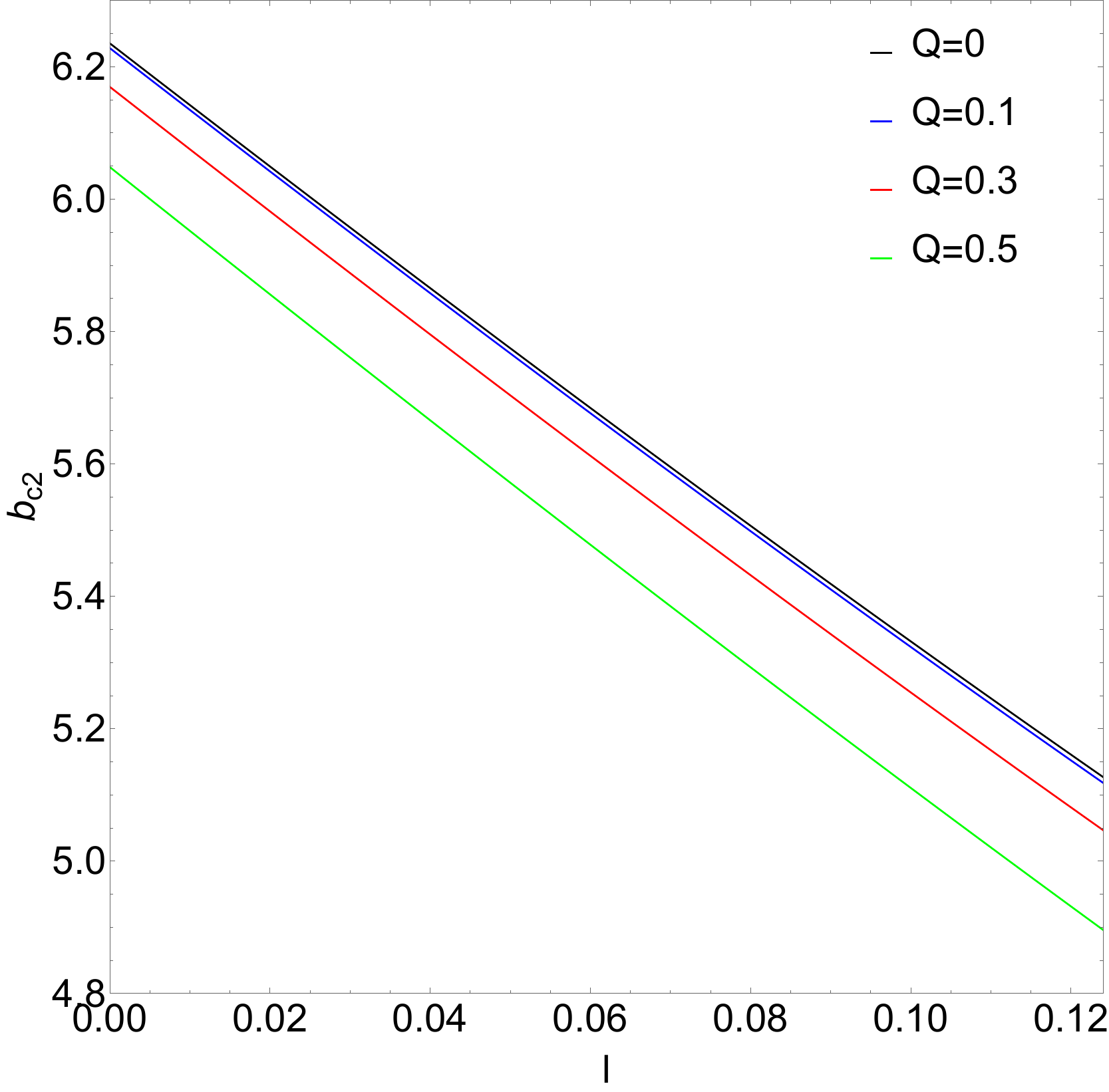}\\
		\end{minipage}
	}
	\caption{(color online) In spacetime $M_2$, the variation of the photon sphere radius $r_{p_{2}}$, impact parameter $b_{c_{2}}$, and the area $S_{p_{2}}$ enclosed by the photon sphere with the Lorentz-violation parameter $l$, is depicted in panels (a), (b), and (c), respectively.}
	\label{S2}
\end{figure*}

\noindent Where $\dot{R}$ denotes the velocity of the moving thin shell. By definition, the metric of spacetime $M_i$ is continuous (\textit{i.e.}, $\mathbf{g}_{ab}^{M_1}(R) = \mathbf{g}_{ab}^{M_2}(R)$) \cite{Nakao:2013hba,Wang:2020emr}. Consequently, the two conserved quantities $E_i = -p_{t_i}$ and $L_i = p_{\phi_i}$ along the geodesic are expressed as

\begin{equation}\label{eqAAA}
    \begin{aligned}
        -\frac{E_2}{\sqrt{F_2(R)}} &= -\frac{\cosh(\theta_1 - \theta_2)}{\sqrt{F_1(R)}} E_1 + \frac{\sinh(\theta_1 - \theta_2)}{\sqrt{F_1(R)}} p_R^{r_1}, &
        L_1 &= L_2.
    \end{aligned}
\end{equation}
Where $p_R^{r_1}$ denotes the value of $p^{r_1}$ on the thin shell. By further combining Eqs.~(\ref{eq7}) and ~(\ref{eqAAA}), the relationship between the impact parameters $b_1$ and $b_2$ for spacetime $M_1$ and spacetime $M_2$ can be derived as follows
\begin{equation}\label{eq11}
\begin{aligned}
\frac{b_1}{b_2} =\sqrt{\frac{F_2(R)}{F_1(R)}} =  \sqrt{\frac{\left(1 + \frac{Q^2}{R^2} - \frac{2 M_2}{R}\right) + \left(1 + \frac{2 Q^2}{R^2}\right) l + \left(1 + \frac{3 Q^2}{R^2}\right) l^2}{\left(1 + \frac{Q^2}{R^2} - \frac{2 M_1}{R}\right) + \left(1 + \frac{2 Q^2}{R^2}\right) l + \left(1 + \frac{3 Q^2}{R^2}\right) l^2}} \equiv  Z.
\end{aligned}
\end{equation}
To study the effects of the charge $Q$ and the Lorentz-violation parameter $l$ on the photon sphere radius $r_{p h_i}$ and the critical collision parameter $b_{c_i}$, calculations were conducted using the formula for the photon sphere radius with $M_1=1$ and $M_2=1.2$. The results are shown in Table.~(\ref{T1}) and ~(\ref{T2}). And the variations of the photon sphere radius and its corresponding area with respect to the charge $Q$ and the Lorentz-violation parameter $l$ are plotted in Figs.~\ref{S1}~and ~\ref{S2}~. Based on the relevant research, the charge $Q$ is limited to the range from $0$ to $0.5$, and the Lorentz-violation parameter $l$ is limited to the range from $0$ to $0.1$. It was observed that when $Q=0$ and $l=0$, the scenario corresponds to the Schwarzschild ATSW, and when $l=0$, it corresponds to the Reissner-Nordstr\"om case. From Figs.~\ref{S1}~,~\ref{S2}~ and Table.~(\ref{T1}),~(\ref{T2}), it is observed that the impact parameter $b_{c_{i}}$, photon sphere radius $r_{p_{i}}$, and the corresponding enclosed area $S_{p_{i}}$ decrease with increasing charge $Q$ and Lorentz-violation parameter $l$.

\begin{table}[h!]\label{T1}
	\centering
	\begin{tabular}{|l|r|c|c|c|c|c|c|c}
		\hline
		$~~Q$ & $0~~~$ & $0.1$ & $0.3$ & $0.5$  \\
		\hline
		$~~b_{c_{1}}$ & $2.97$ & $2.96319$ & $2.90474$ & $2.78891$  \\
		\hline
		$~~r_{p_{1}}$ & $5.11841$ & $5.1096$ & $5.0378$ & $4.8863$ \\
		\hline
		$~~b_{c_{2}}$ & $3.564$ & $3.55833$ & $3.51224$ & $3.41614$\\
		\hline
		$~~r_{p_{2}}$ & $6.1421$ & $6.13476$ & $6.07531$ & $5.95206$ \\
		\hline
	\end{tabular}
	\caption{The variation of photon sphere radius $r_{p_{i}}$ and critical impact parameter $b_{c_{i}}$ with respect to charge $Q$, in KR spacetime with Lorentz violation parameter $l=0.01$. The charge values are taken as $Q=$ $0,\,0.1,\,0.3,\,0.5$. The discussion covers two spacetime configurations with mass parameters $M_1=1$ and $M_2=1.2$ respectively.}\label{T1}
\end{table}

\begin{table}[h!]\label{T2}
	\centering
	\begin{tabular}{|l|r|c|c|c|c|c|c|c}
		\hline
		$~~l$ & $0~~~$ & $0.01$ & $0.05$ & $0.1$  \\
		\hline
		$~~b_{c_{1}}$ & $2.93875$ & $2.90747$ & $2.78228$ & $2.62677$  \\
		\hline
		$~~r_{p_{1}}$ & $5.11679$ & $5.0378$ & $4.72632$ & $4.34993$ \\
		\hline
		$~~b_{c_{2}}$ & $3.54929$ & $3.51224$ & $3.36413$ & $3.18053$\\
		\hline
		$~~r_{p_{2}}$ & $6.16962$ & $6.07531$ & $5.70357$ & $5.25473$ \\
		\hline
	\end{tabular}
	\caption{The variation of the photon sphere radius $r_{p_{i}}$ and the critical impact parameter $b_{c_{i}}$ with respect to the Lorentz-violation parameter $l$, in the KR spacetime with ATSW (charge $Q=0.3$). The values of the Lorentz-violation parameter $l$ are taken as $l=$ $0,\,0.01,\,0.05,\,0.1$. The analysis is conducted for two distinct spacetime configurations with mass parameters $M_1=1$ and $M_2=1.2$, respectively.}\label{T2}
\end{table}
\begin{figure*}[htbp]
	\centering
	\subfigure[The effective potential $V_{e\!f\!f}$ of a BH in the KR spacetime varies with the charge $Q$.]{
		\begin{minipage}[t]{0.33\linewidth}
			\includegraphics[width=2.4in]{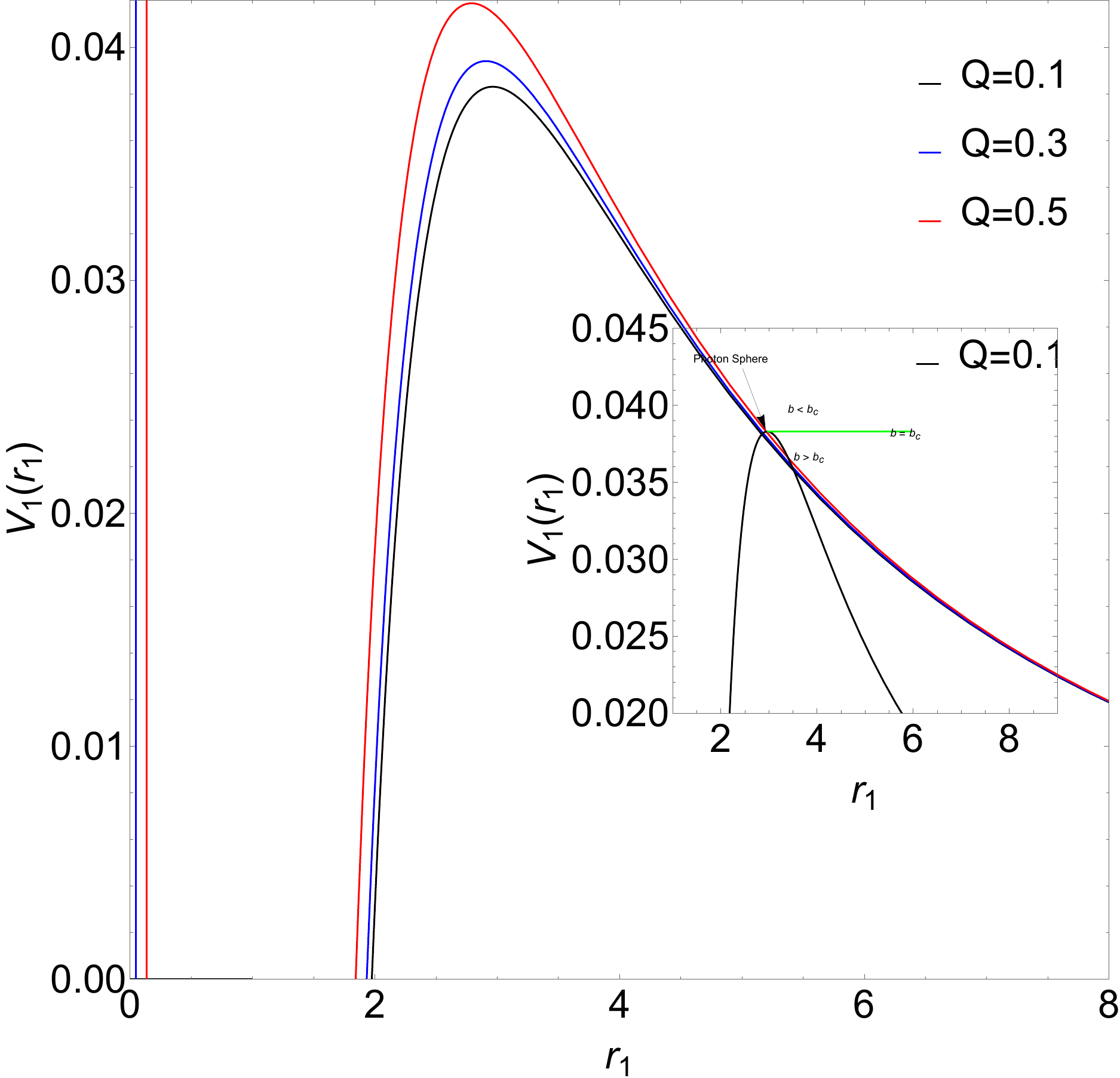}\\
		\end{minipage}
	}
	\subfigure[The effective potential $V_{e\!f\!f}$ of the ATSW in the KR spacetime varies with the charge $Q$.]{
		\begin{minipage}[t]{0.33\linewidth}
			\includegraphics[width=2.4in]{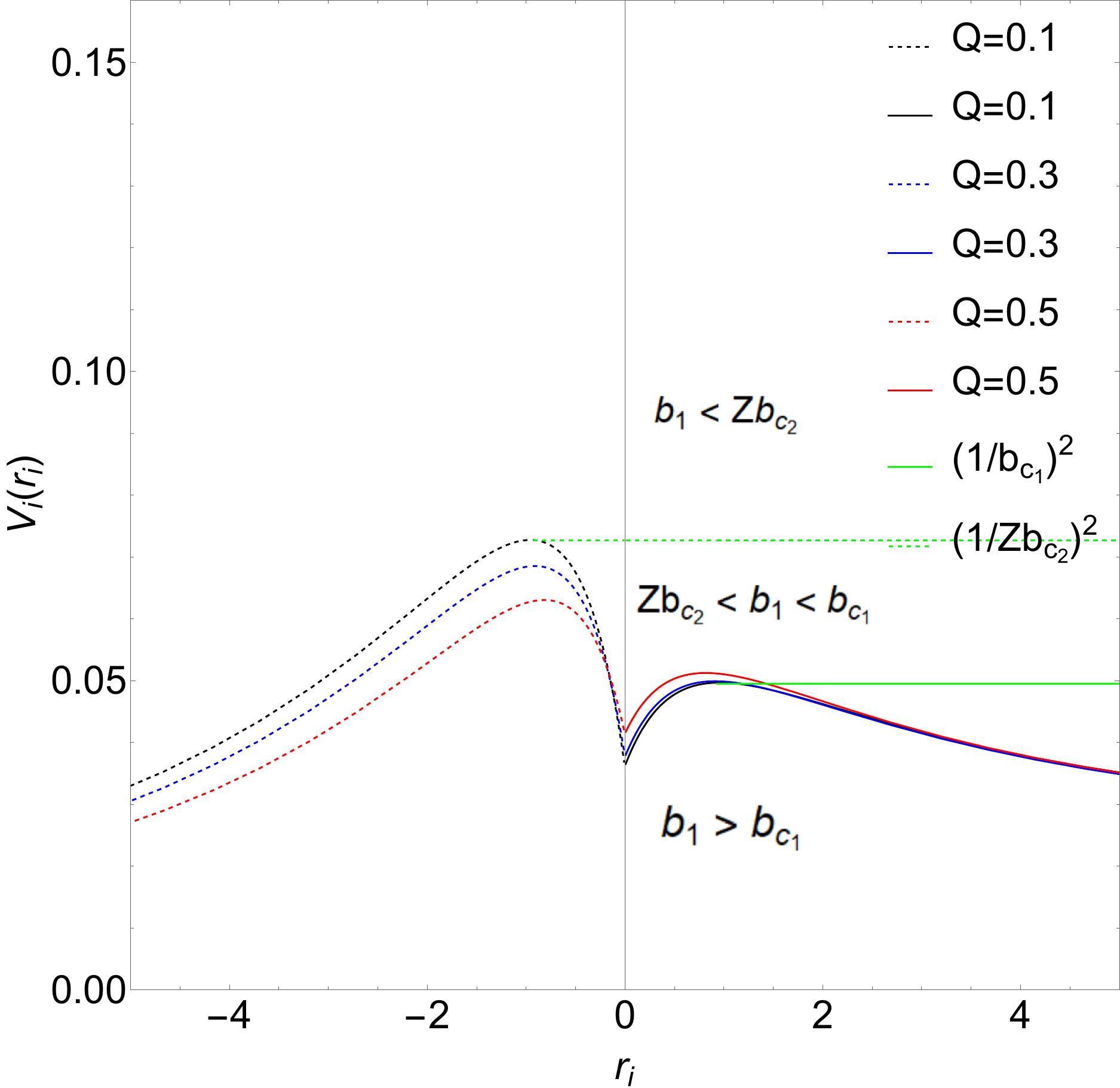}\\
		\end{minipage}
	}
	\caption{(color online) With Lorentz-violation parameter $l=0.01$, we analyze the effective potential of the BH  and ATSW in the KR field as functions of $Q$. The left plot shows the effective potential for BH, while the right plot displays that for ATSW. Various $Q$ values are considered, specifically $Q=$ $0,\,0.1,\,0.3,\,0.5$, respectively. In the right plot, the solid and dashed lines correspond to the effective potentials for$M_1$ and $M_2$, respectively, with parameters set as $M_1=1$, $M_2=1.2$, and $R=2.6$.}
	\label{A1}
\end{figure*}

\begin{figure*}[htbp]
	\centering
	\subfigure[The effective potential $V_{e\!f\!f}$ of BH in the KR field under varying Lorentz-violation parameter $l$.]{
		\begin{minipage}[t]{0.33\linewidth}
			\includegraphics[width=2.4in]{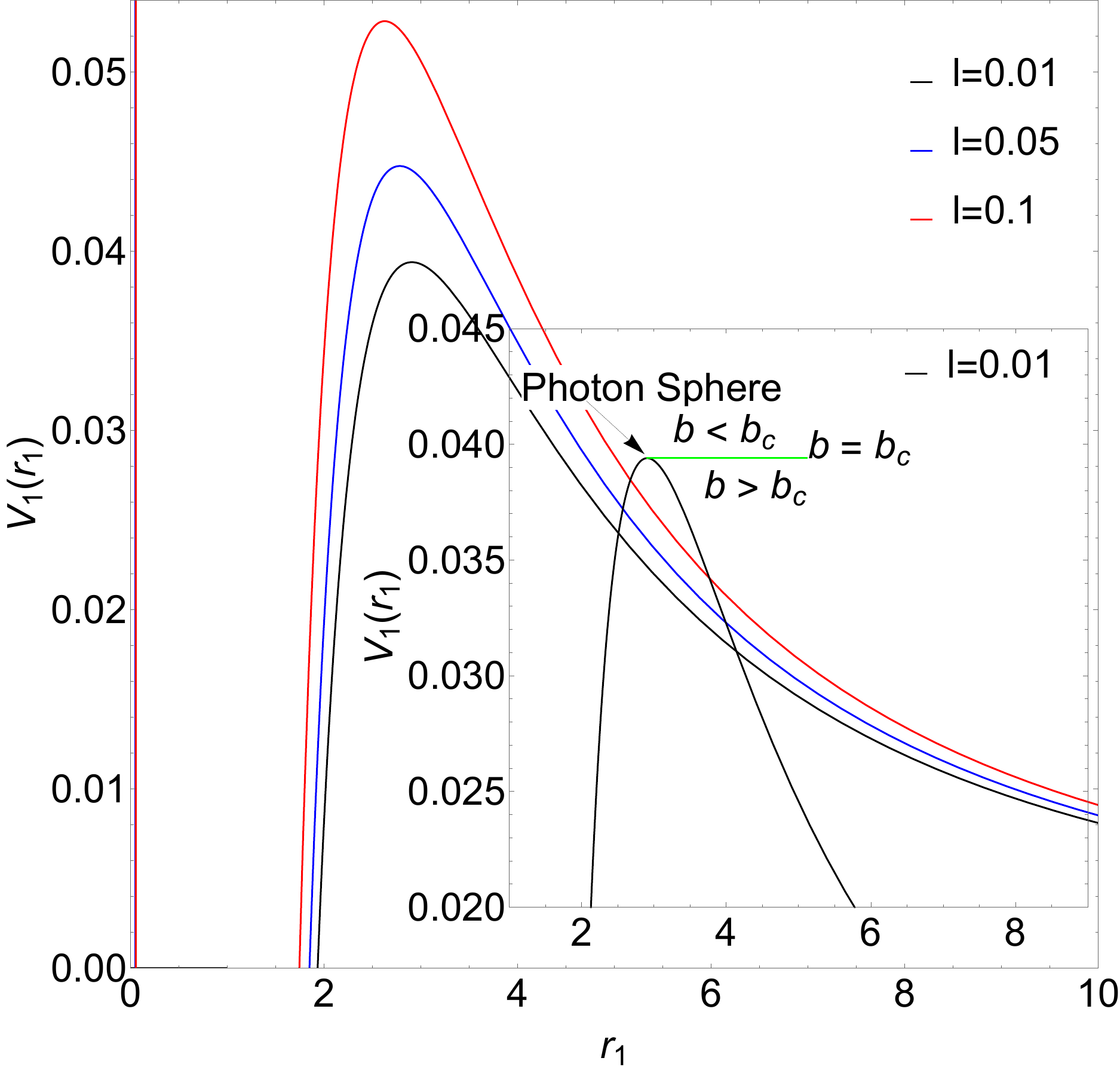}\\
		\end{minipage}
	}
	\subfigure[The effective potential  $V_{e\!f\!f}$ energy of ATSW in the KR field varies with the change of the Lorentz-violation parameter $l$.]{
		\begin{minipage}[t]{0.33\linewidth}
			\includegraphics[width=2.4in]{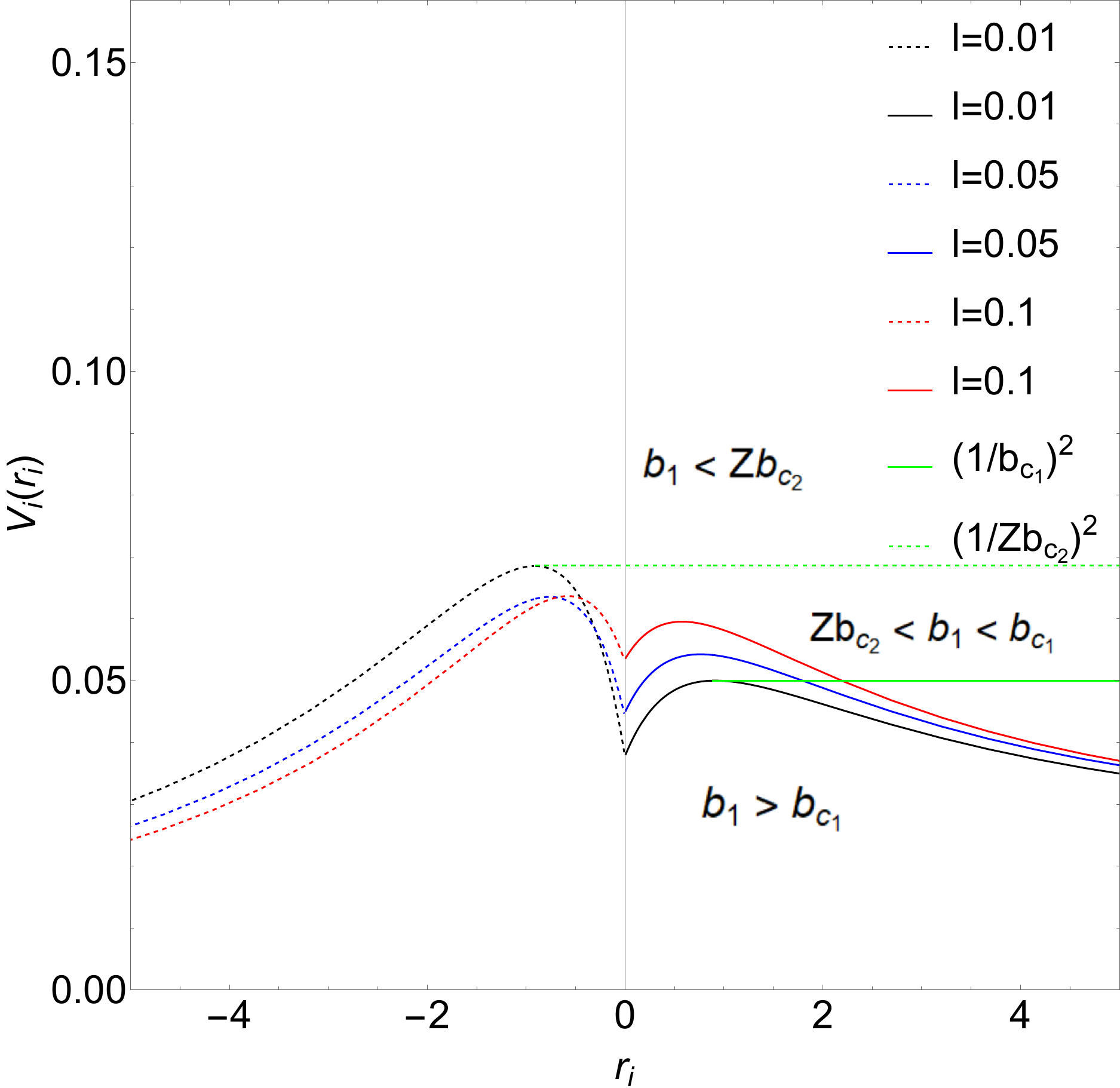}\\
		\end{minipage}
	}
	\caption{(color online)The effective potential energy of BH (with charge $Q=0.3$) and ATSW (with charge $Q=0.3$) in the KR field varies with the value of $l$. On the left, the effective potential energy of BH is depicted, while on the right, that of ATSW is shown. We have designed different $Q$ values and set $l$ to $0$, $0.01$, $0.05$, and $0.1$. In the right figure, the solid and dashed lines represent the effective potential energies of $M_1$ and $M_2$, respectively. Specifically, we have set $M_1$,$M_2$, and $R$ to $1$, $1.2$, and $2.6$, respectively.}
	\label{A2}
\end{figure*}

The effective potential energies of BH and ATSW in the KR field are depicted in Figs. ~\ref{A1}~ and  ~\ref{A2}~, respectively. By examining the effective potential energy of BH (with charge $Q=0.3$) in Figs.~\ref{A1}~(a) and ~\ref{A2}~(a), we discern three distinct photon trajectories as illustrated in the diagram on the lower right side. For the scenario where $l=0.01$ and $Q=0.1$, as shown by the black line in Fig.~\ref{A1}~, when $b_1>b_{c_1}=5.11841$, the gravitational influence results in the deflection of the photon trajectory. When $b_1=b_{c_1}$, the photon will perpetually orbit around the BH in a circular path. When $b_1<b_{c_1}$, due to the intense gravitational force, the photon will directly plunge into the BH. Furthermore, it is noted that with the increase in charge $Q$ and the Lorentz-violation parameter $l$, the effective potential energy of the BH also rises, and the position corresponding to the peak shifts inward.

In Figs.~\ref{A1}~(b) and ~\ref{A2}~(b), we plot the effective potential of the ATSW as a function of the Lorentz-violation parameter $l$ for a fixed charge $Q=0.1$. The dashed line on the left corresponds to spacetime $M_2$, while the solid line on the right corresponds to spacetime $M_1$. The effective potential of spacetime $M_2$ is constrained by a factor of $Z^2$.
From the observations of Figs.~\ref{A1}~(b) ~\ref{A2}~(b), we note that, for a given charge $Q$ or Lorentz-violation parameter $l$, photons exhibit three distinct trajectories~\cite{Gralla:2019xty}. For the case where $l=0.1$ and $Q=0.1$, as indicated by the black line in Fig.~\ref{A1}~(b): When $b_1<Zb_{c_1}=3.708$, the photon falls from spacetime $M_1$ into spacetime $M_2$ and moves toward infinity.When $Zb_{c_2}<b_1<b_{c_1}=5.11841$, the photon enters the potential barrier in spacetime $M_2$ and then returns to spacetime $M_1$.
When $b_1<Zb_{c_2}$, the photon moves to the potential barrier in spacetime $M_1$ and then travels toward infinity in $M_1$.

\begin{figure*}[htbp]
	\centering
	\subfigure[$Q=0.1$]{
		\begin{minipage}[t]{0.28\linewidth}
			\includegraphics[width=2.1in]{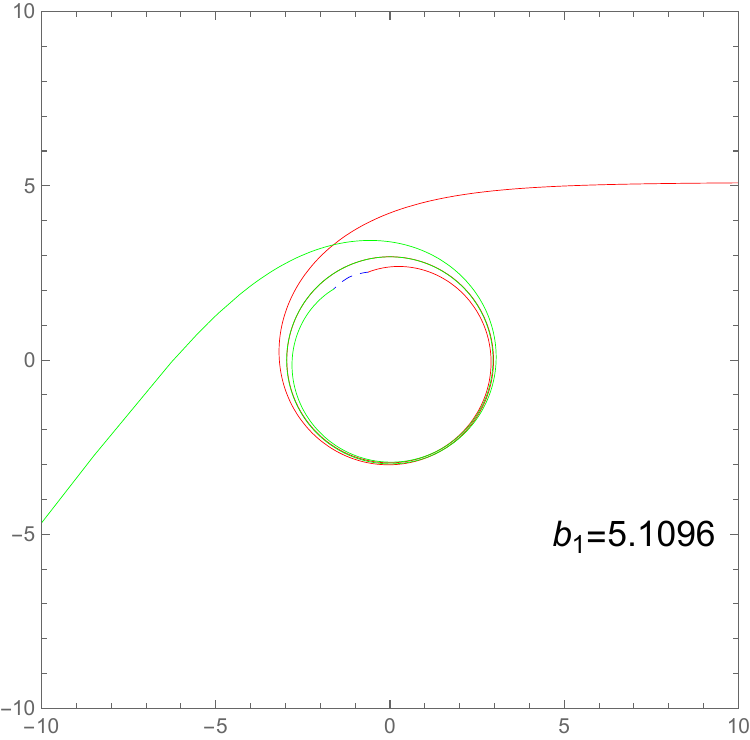}\\
			\includegraphics[width=2.1in]{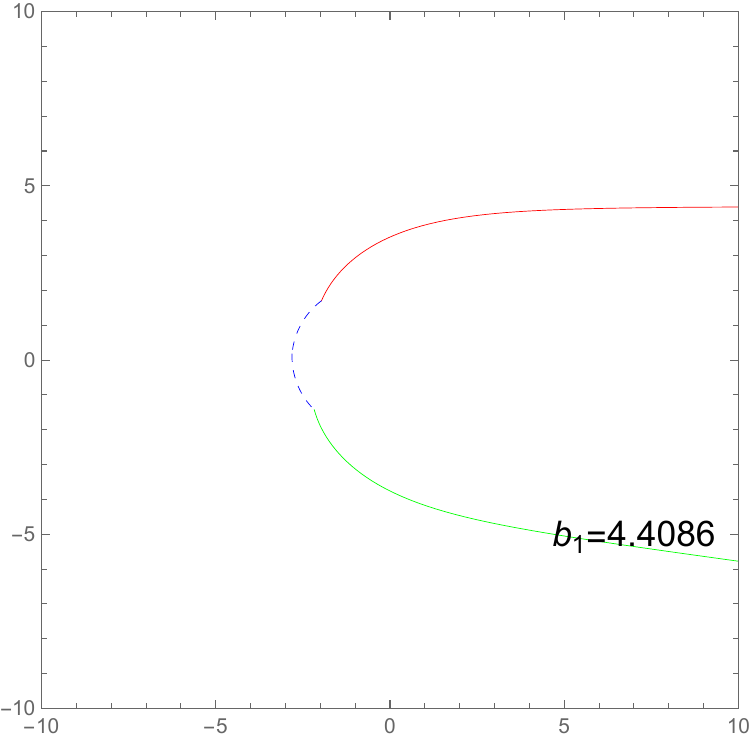}\\
			\includegraphics[width=2.1in]{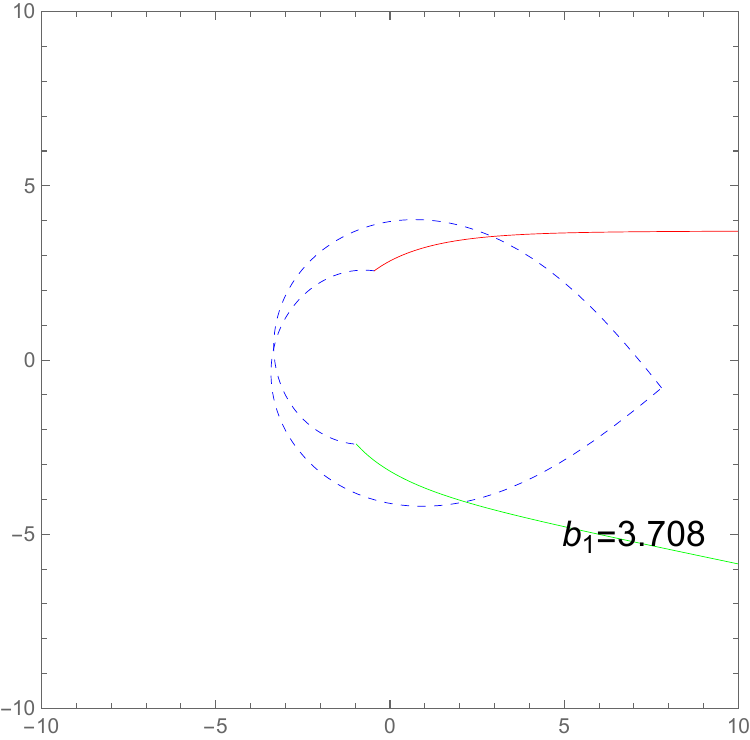}\\
		\end{minipage}
	}
	\subfigure[$Q=0.3$]{
		\begin{minipage}[t]{0.28\linewidth}
			\includegraphics[width=2.1in]{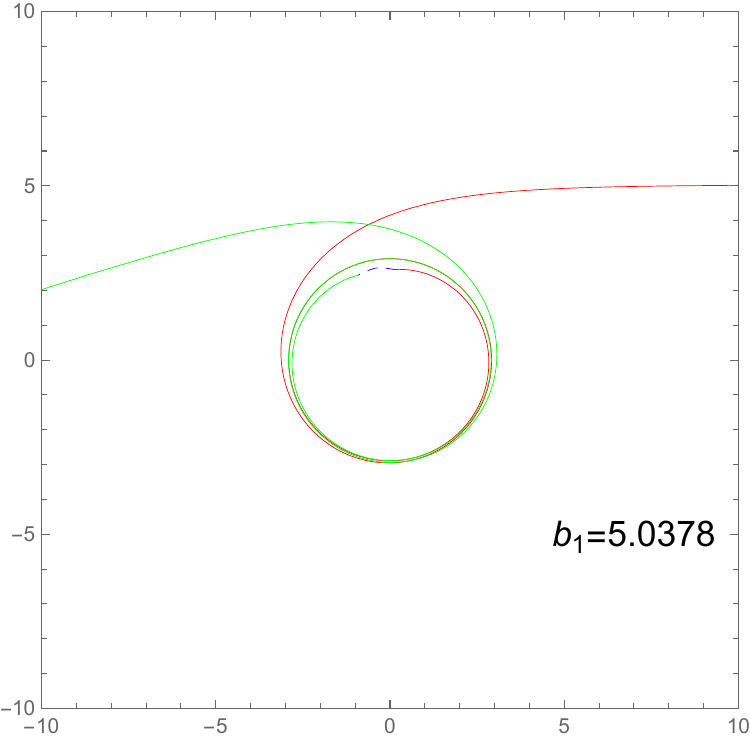}\\
			\includegraphics[width=2.1in]{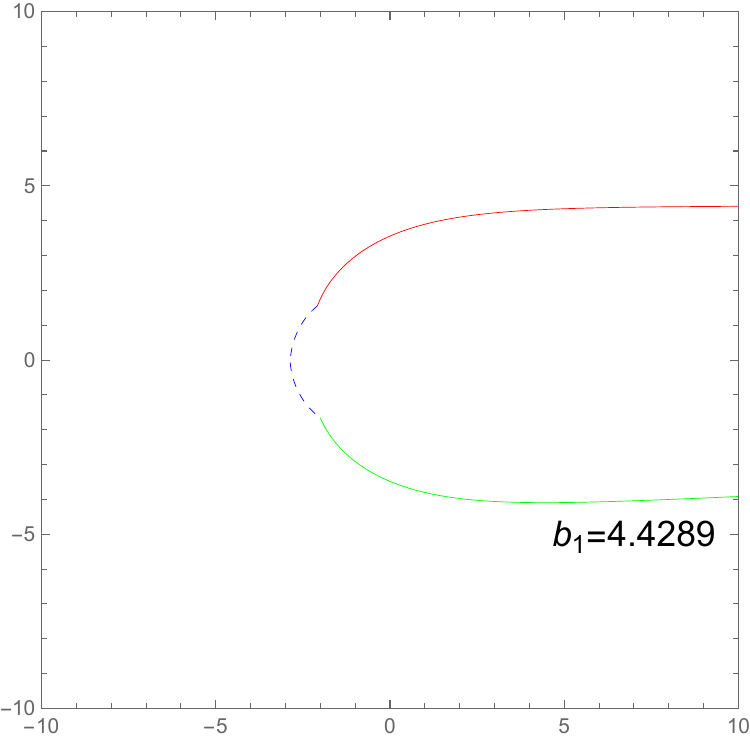}\\
			\includegraphics[width=2.1in]{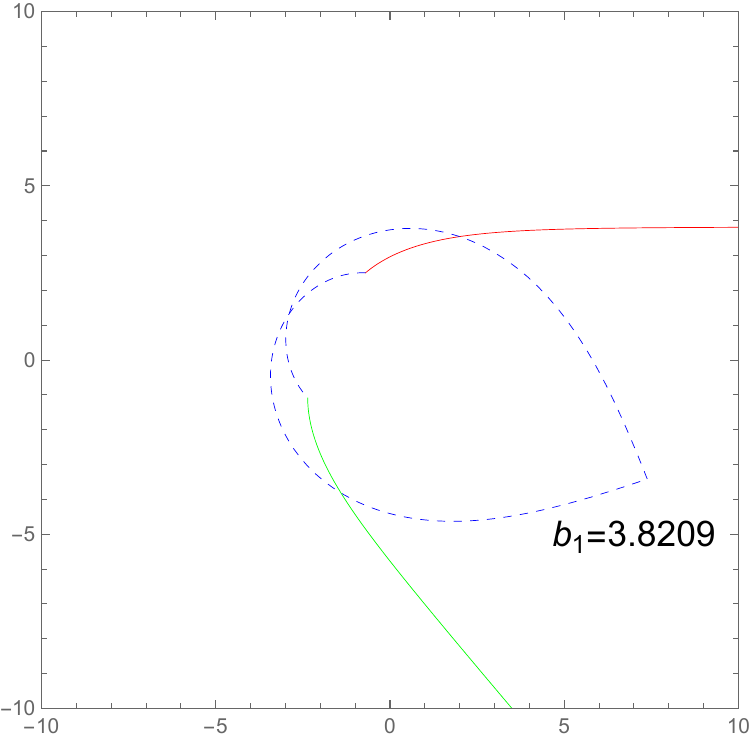}\\
		\end{minipage}
	}
	\subfigure[$Q=0.5$]{
		\begin{minipage}[t]{0.28\linewidth}
			\includegraphics[width=2.1in]{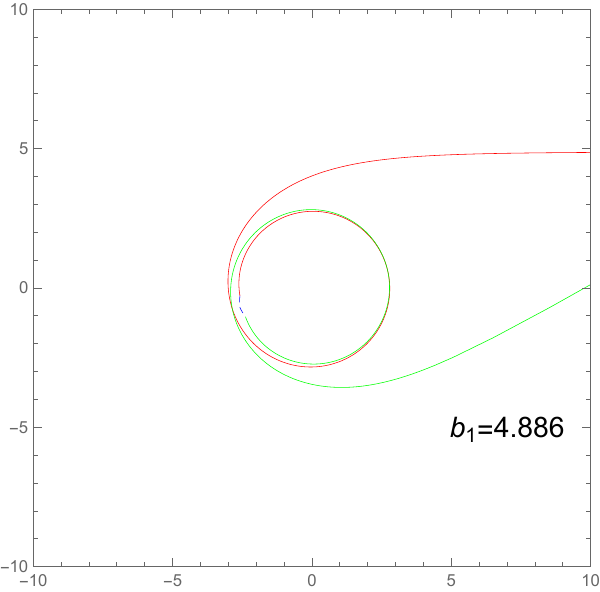}\\
			\includegraphics[width=2.1in]{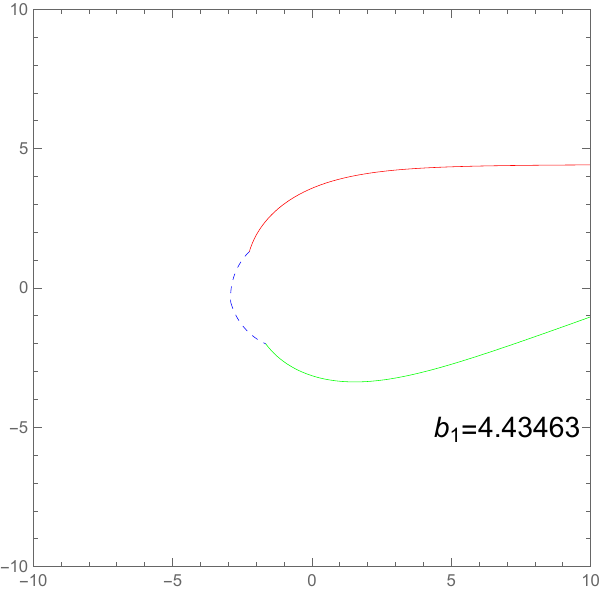}\\
			\includegraphics[width=2.1in]{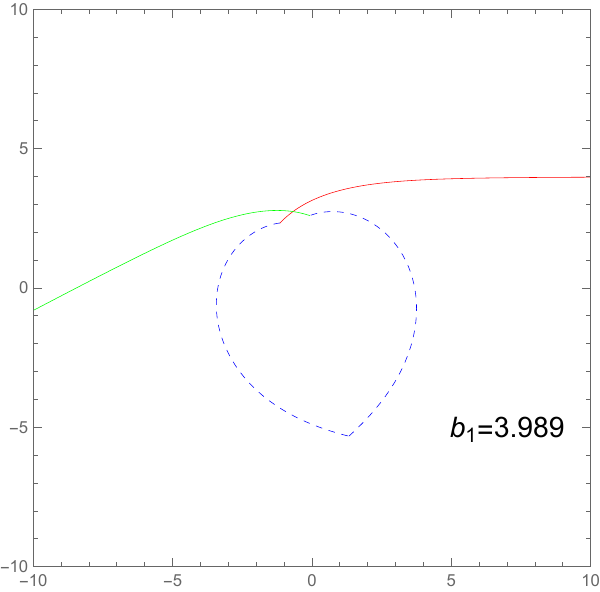}\\
		\end{minipage}
	}
	\caption{(color online) The influence of the charge $Q$ on photon trajectories in polar coordinates $\left(r_1, \varphi\right)$, specifically for impact parameters in the range $Zb_{c_2}<b_1<b_{c_1}$. The first, second, and third rows correspond to the cases of $Q = 0.1$, $Q = 0.3$, and $Q = 0.5$, respectively. The trajectories are denoted as follows: the red lines represent incident photons in spacetime $M_1$, the green lines represent outgoing photons in spacetime $M_1$, and the blue dashed lines depict the photon trajectories within spacetime $M_2$. The parameters are fixed at $l = 0.01$, $R = 2.6$, $M_2=1$, and $M_2=1.2$.}
	\label{A3}
\end{figure*}

\begin{figure*}[htbp]
	\centering
	\subfigure[$l=0.01$]{
		\begin{minipage}[t]{0.28\linewidth}
			\includegraphics[width=2.1in]{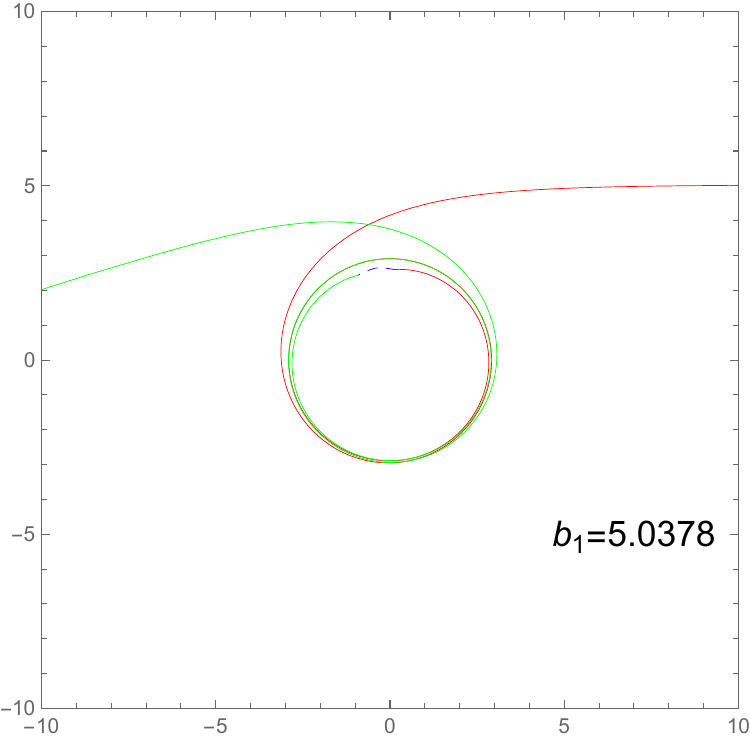}\\
			\includegraphics[width=2.1in]{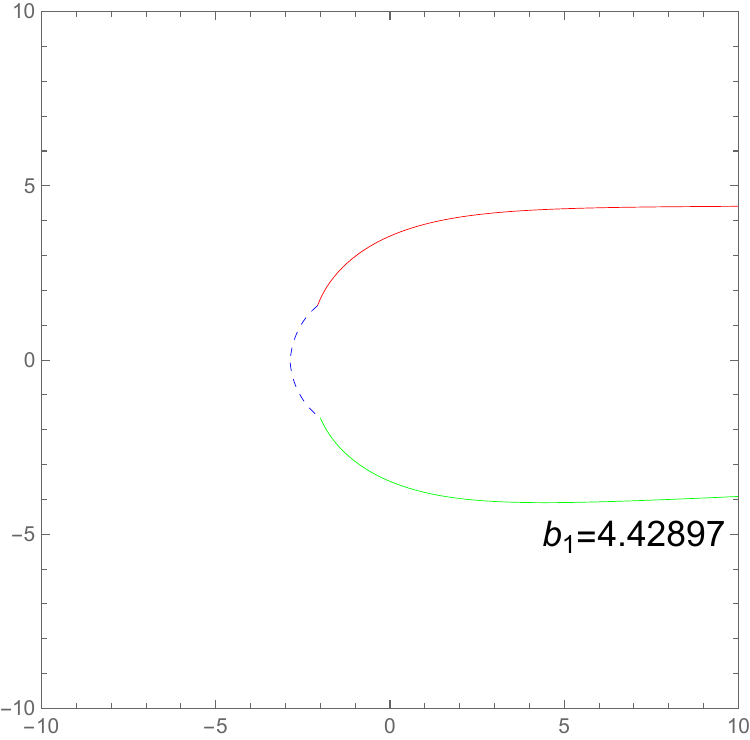}\\
			\includegraphics[width=2.1in]{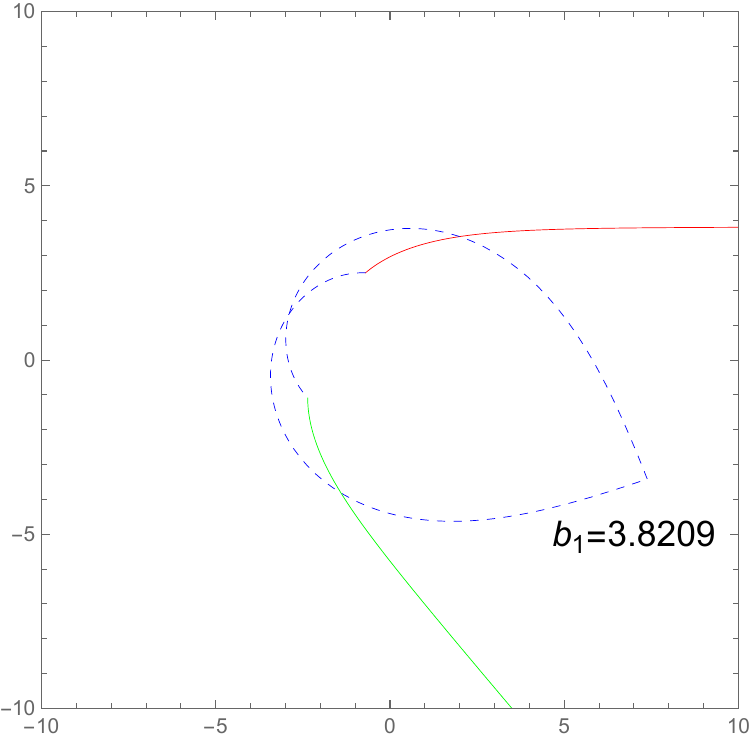}\\
		\end{minipage}
	}
	\subfigure[$l=0.05$]{
		\begin{minipage}[t]{0.28\linewidth}
			\includegraphics[width=2.1in]{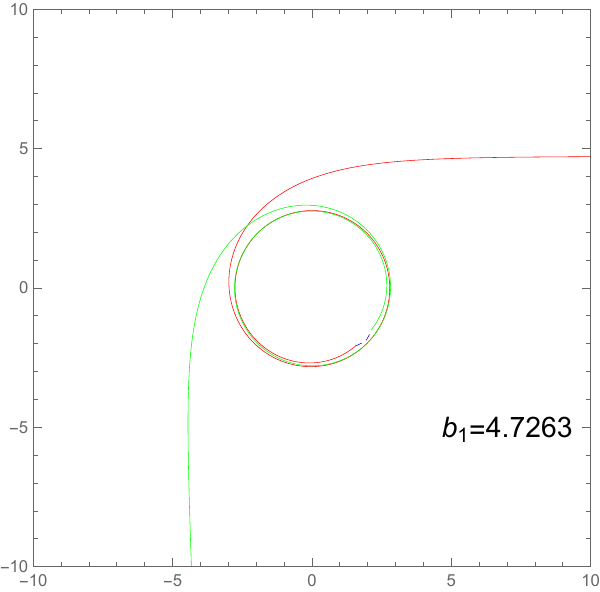}\\
			\includegraphics[width=2.1in]{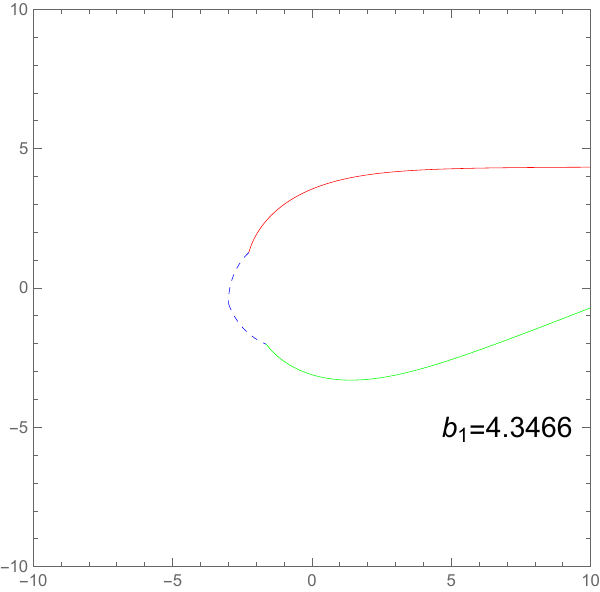}\\
			\includegraphics[width=2.1in]{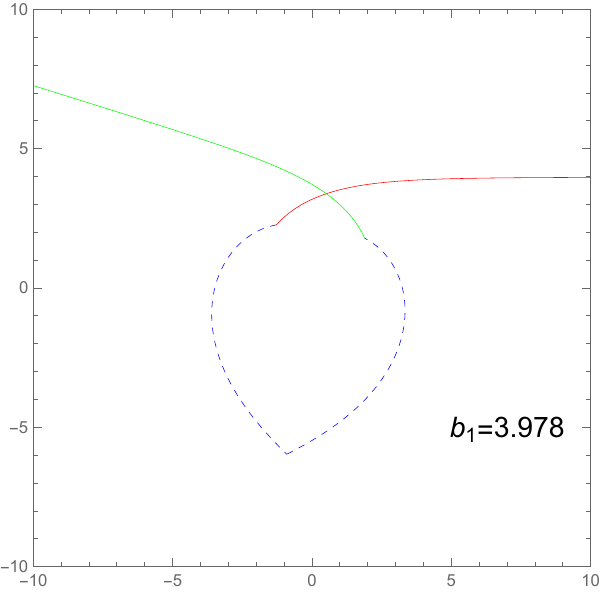}\\
		\end{minipage}
	}
	\subfigure[$l=0.1$]{
		\begin{minipage}[t]{0.28\linewidth}
			\includegraphics[width=2.1in]{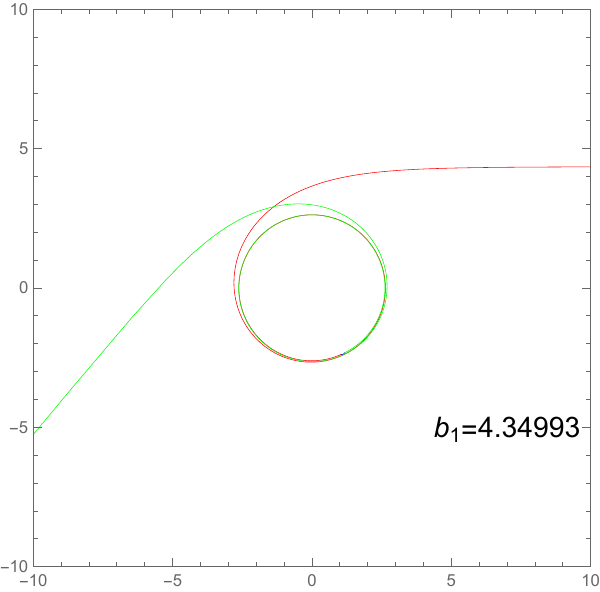}\\
			\includegraphics[width=2.1in]{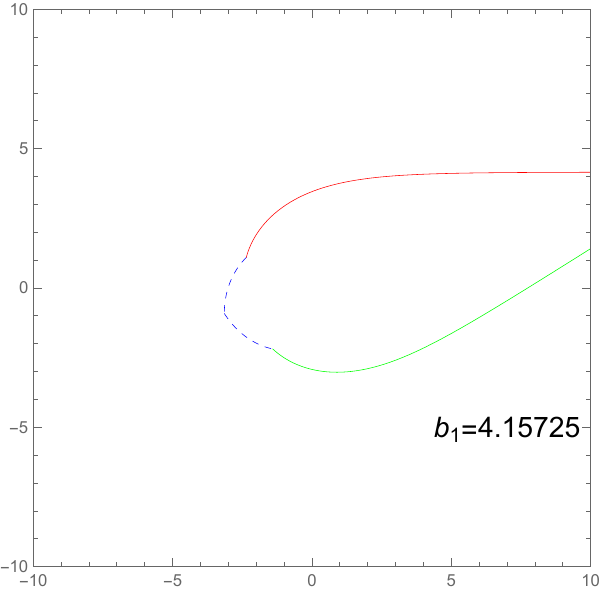}\\
			\includegraphics[width=2.1in]{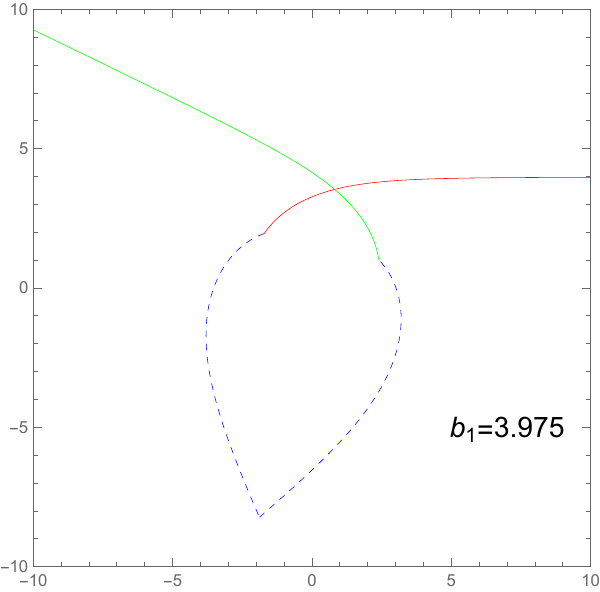}\\
		\end{minipage}
	}
	\caption{(color online) Plotted in polar coordinates $\left(r_1, \varphi\right)$, this figure illustrates the dependence of photon trajectories in spacetime $M_i$ on the Lorentz-violation parameter $l$, for impact parameters in the range $Zb_{c_2}<b_1<b_{c_1}$. The first, second, and third rows correspond to $l=0.01$, $l=0.05$, and $l=0.1$, respectively. The red curves represent incident photons in spacetime $M_1$, the green curves represent outgoing photons in spacetime $M_1$, and the blue dashed curves depict trajectories within spacetime$M_2$. The parameters are fixed at $Q=0.3$, $R =2.6$, $M_1= 1$, and $M_2 = 1.2$.}
	\label{A4}
\end{figure*}

To better understand the optical characteristics of the ATSW in the KR field, we calculated the azimuthal change and trajectory of photons in spacetime $M_i$, and derived the photon trajectory equation based on Eq.~(\ref{eq6})
\begin{equation}\label{eq12}
\frac{1}{b_i^2}-\frac{f_i\left(r_i\right)}{r_i^2}=\frac{1}{r_i^4}\left(\frac{\mathrm{~d} r_i}{\mathrm{~d} \phi}\right)^2.
\end{equation}
To simplify the calculation process, we introduced the substitution 
$x=1 / r$ and derived a new photon trajectory equation
\begin{equation}\label{eq13}
G_i\left(x_i\right)=\left(\frac{\mathrm{d} x_i}{\mathrm{~d} \varphi}\right)^2=\frac{1}{b_i^2}-x_i^2\left[\left(1+{Q^2 x_i^2}-{2 M_i x_i}\right)+\left(1+{2 Q^2 x_i^2}\right) l+\left(1+{3 Q^2 x_i^2}\right) l^2 \right].
\end{equation}
When $b_1>b_{c_1}=5.11841$, the photon moves to the potential barrier in spacetime $M_1$ and subsequently remains within $M_1$. The position of this barrier is determined by the smallest positive root, $x_{1 \text { min }}$, of the equation $G_1\left(x_1\right)=0$. The photon's azimuthal change is then calculated using formula
\begin{equation}\label{eq14}
\varphi_1\left(b_1\right)=2 \int_0^{x_1^{\min }} \frac{\mathrm{d} x_1}{\sqrt{G_1\left(x_1\right)}}, \quad b_1>b_{c_1}.
\end{equation}
When $Zb_{c_2}<b_1<b_{c_1}=5.11841$, the photon travels from spacetime $M_1$ through the throat into spacetime $M_2$. Upon reaching the potential barrier in $M_2$-whose position is defined by the largest positive root, 
$x_{2 \text { min }}$, of the equation $G_2\left(x_2\right)=0$, it is reflected back, returning to spacetime $M_1$. The total azimuthal change of the photon is then determined by summing its angular displacements in both spacetimes, $M_1$ and $M_2$
\begin{equation}\label{eq15}
\varphi_1^*\left(b_1\right)=\int_0^{1 / R} \frac{d x_1}{\sqrt{G_1\left(x_1\right)}}, \quad b_1<b_{c_1}.
\end{equation}
When $b_1<Zb_{c_2}$, the photon passes through the throat from spacetime $M_1$ into spacetime $M_2$ and propagates to infinity. The azimuthal change of the photon is then solely determined by its trajectory within spacetime $M_2$
\begin{equation}\label{eq16}
\varphi_2\left(b_2\right)=2 \int_{x_2^{\max }}^{1 / R} \frac{d x_2}{\sqrt{G_2\left(x_2\right)}}, \quad b_2>b_{c_2}.
\end{equation}
With Eqs.~(\ref{eq11})~(\ref{eq12})~(\ref{eq13}), we can plot the photon trajectories for different values of the charge $Q$ and the Lorentz-violation parameter $l$, as shown in Figs.~\ref{A3}~ and ~\ref{A4}~. Taking $Q=0.1$ and $l=0.01$ as an example, Fig. ~\ref{A3}~(a), (b), and (c) illustrate three typical types of photon motion. For the case where $Zb_{c_2}<b_1<b_{c_1}$, photons originating from infinity in spacetime $M_1$ exhibit an increasing number of trajectory loops in spacetime $M_2$ as the impact parameter $b_i$ decreases. By comparing the corresponding columns in Figs.~\ref{A3}~ and ~\ref{A4}~, we observe that $b_{c_1}$ decreases with an increase in either $Q$ or $l$, while $Zb_{c_2}$ shows the opposite trend. Furthermore, it is evident that the photon trajectories are more significantly influenced by the Lorentz-violation parameter $l$.

\section{The Optical Appearance of an ATSW}\label{sec3}

The photon reflection mechanism in an ATSW differs from that of a BH. To investigate the differences in observable appearances between the ATSW in the KR field and a BH, we will consider incident photons with specific impact parameters. In this section, we study the ATSW in the KR field surrounded by an optically and geometrically thin accretion disk. By employing two radiation models, we compare the observational signatures of the ATSW and the BH.

\subsection{The trajectory of a photon}
First, in a stationary coordinate system with the BH fixed in the equatorial plane, consider a static observer at infinite distance in the direction of the North Pole. We assume that the light source surrounding the black hole is a geometrically and optically thin accretion disk situated on the equatorial plane. Based on the backward ray-tracing approach, if the light rays emitted by the observer intersect the accretion disk, the light emanating from the corresponding position on the disk can be detected by the observer. The total number of orbits, as a function of $b$, which can be expressed as

\begin{equation}\label{eq17}
n=\varphi / 2 \pi.
\end{equation}

From the perspective of an observer at infinity, photon trajectories around a black hole are classified into three distinct types according to their orbit number $n$:

$a$. Direct Emission is defined as trajectories with $n < 0.75$, characterized by a single intersection with the equatorial plane.

$b$. Lensing Ring refers to trajectories where $0.75 < n < 1.25$, which intersect the equatorial plane exactly twice.

$c$. Photon Ring comprises trajectories with $n > 1.25$, which cross the equatorial plane a minimum of three times.

\begin{figure*}[htbp]
	\centering
	\subfigure[Orbit number $n_1$.]{
		\begin{minipage}[t]{0.33\linewidth}
			\includegraphics[width=2.4in]{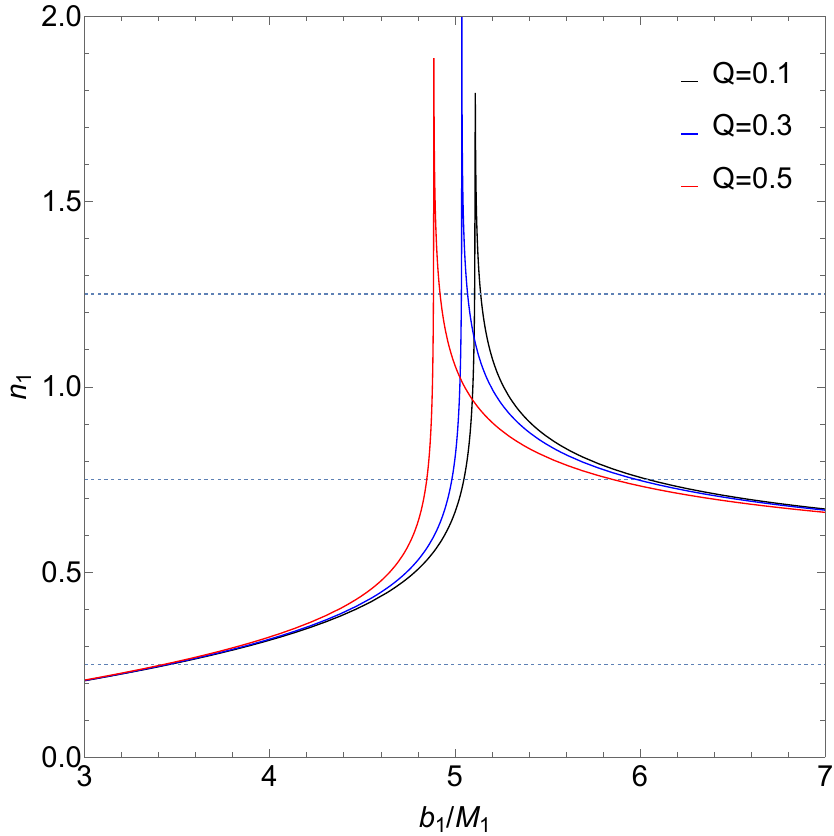}\\
		\end{minipage}
	}
	\subfigure[Orbit number $n_2$, $n_3$.]{
		\begin{minipage}[t]{0.33\linewidth}
			\includegraphics[width=2.4in]{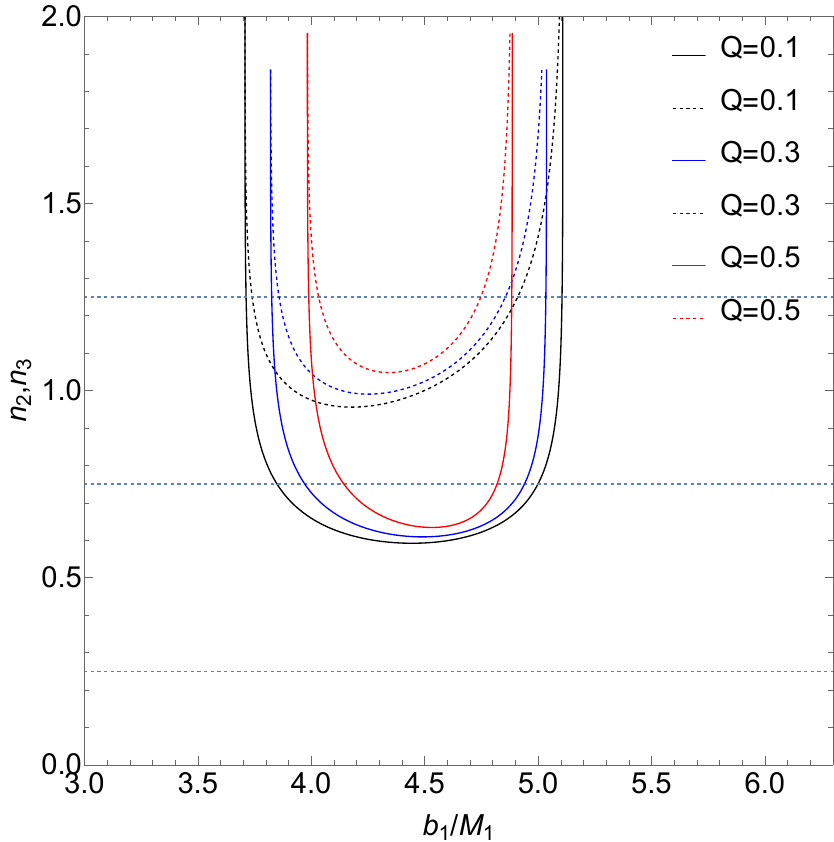}\\
		\end{minipage}
	}
	\caption{(color online)A diagram illustrating the relationship between the number of photon trajectories $n$ and the impact parameter $b_1$ around an ATSW (with $l = 0.01$) in a KR field is presented. The left panel displays the trajectory count $n_1$, while the right panel shows the counts for $n_2$ (solid lines) and $n_3$ (dashed lines). The results for different charge values are compared: $Q = 0.1$ (black lines), $Q = 0.3$ (blue lines), and $Q = 0.5$ (red lines). The parameters are fixed at $R = 2.6$, $M_1 = 1$, and $M_2 = 1.2$.}
	\label{A5}
\end{figure*}

\begin{figure*}[htbp]
	\centering
	\subfigure[Orbit number $n_1$.]{
		\begin{minipage}[t]{0.33\linewidth}
			\includegraphics[width=2.4in]{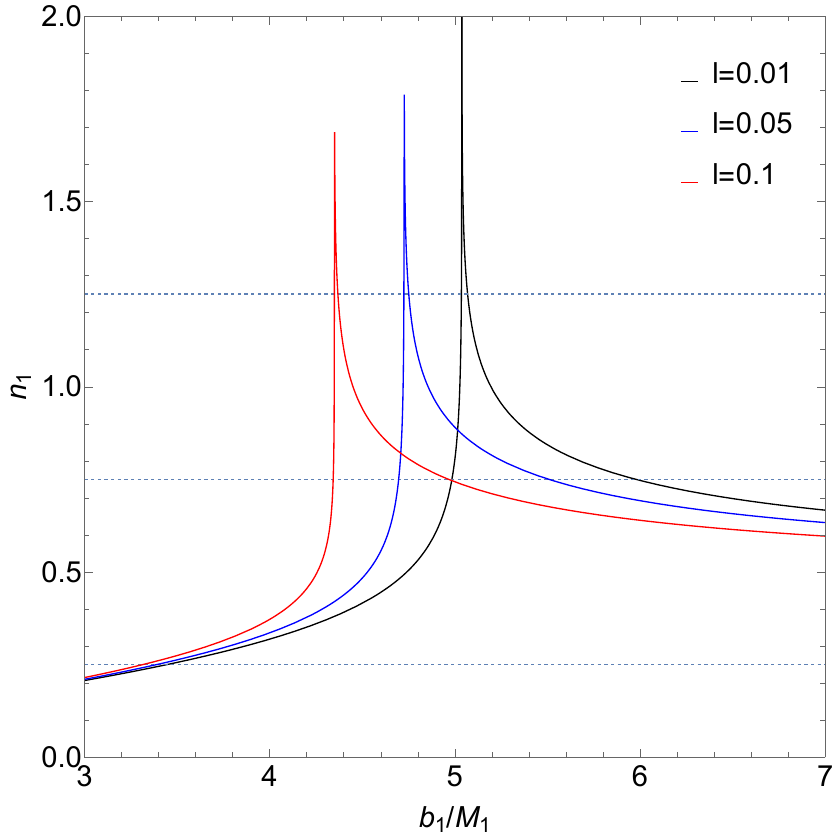}\\
		\end{minipage}
	}
	\subfigure[Orbit number $n_2$, $n_3$.]{
		\begin{minipage}[t]{0.33\linewidth}
			\includegraphics[width=2.4in]{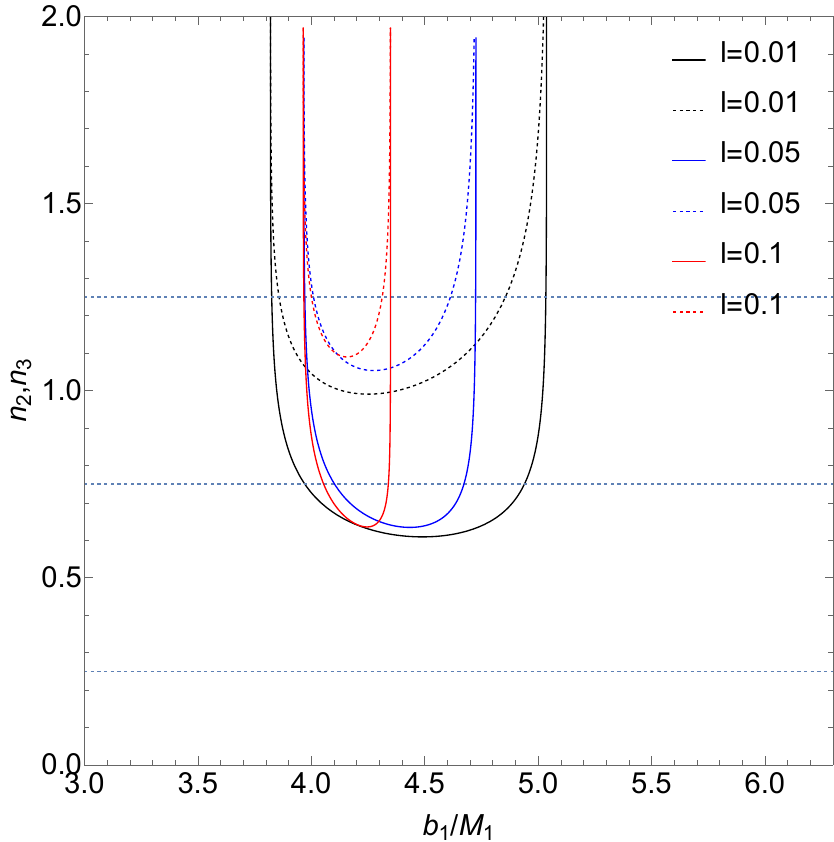}\\
		\end{minipage}
	}
	\caption{(color online) In the KR field, we present the relationship between the number of photon trajectories $n$ with different $l$-values and the collision parameter $b_1$ around the ATSW (with $Q=0.3$). The left figure shows the trajectory count for $n_1$, while the right figure illustrates the trajectory counts for $n_2$ (solid lines) and $n_3$ (dashed lines). We analyze the trajectory counts for photons under different Lorentz-violation parameters $l$: $l=0.01$ (black lines), $l=0.05$ (blue lines), and $l=0.1$ (red lines). The parameters are set as $R=2.6$, $M_1 = 1$, and $M_2 = 1.2$.}
	\label{A6}
\end{figure*}

The trajectory is defined as follows for an ATSW: photons originating from spacetime $M_1$ with an impact parameter $Zb_{c_2}<b_1<b_{c_1}$ first enter the throat and then pass into spacetime $M_2$
\begin{equation}\label{eq18}
n_1\left(b_1\right)=\frac{\varphi_1\left(b_1\right)}{2 \pi},
\end{equation}
\begin{equation}\label{eq19}
n_2\left(b_2\right)=\frac{\varphi_1^*\left(b_1\right)+\varphi_2\left(b_1 / Z\right)}{2 \pi},
\end{equation}
\begin{equation}\label{eq20}
n_3\left(b_1\right)=\frac{2 \varphi_1^*\left(b_1\right)+\varphi_2\left(b_1 / Z\right)}{2 \pi}.
\end{equation}
Here, $n_2$ and $n_3$ serve as additional orbit functions for the ATSW, corresponding to additional photons. This situation is analogous to that of a BH. To better illustrate the relationship between the number of photon trajectories and the impact parameter, we plotted Figs.~\ref{A5}~ and ~\ref{A6}~using Eqs.~(\ref{eq18})~(\ref{eq19}) and~(\ref{eq20}).

Figs.~\ref{A5}~ and ~\ref{A6}~ illustrate the relationship between the number of photon trajectories and the impact parameter $b_1$ around an ATSW, under the influence of charge $Q$ and the Lorentz-violation parameter $l$. In Figs.~\ref{A5}~(a) and ~\ref{A6}~(a), it is observed that the photon trajectories for $n_1$ are similar to those of a black hole, indicating that these photons will continue to propagate in spacetime $M_1$. In contrast, trajectories corresponding to $n_2$ and $n_3$ appear only in the ATSW. When a photon enters the throat from spacetime $M_1$, passes through the potential barrier in spacetime $M_2$, and returns to spacetime $M_1$, the outgoing photons intersect with the back side of the thin accretion disk if the conditions in Figs.~\ref{A5}~(b) and ~\ref{A6}~(b) are met—specifically, when $n_2<0.75$ (solid line) and $0.75<n_3$ (dashed line). Conversely, when $n_2<1.25$ and $1.25<n_3$, the outgoing photons intersect with the front side of the thin accretion disk. Furthermore, analysis of Figs.~\ref{A5}~(b) and ~\ref{A6}~(b) indicates that the range of the impact parameter $b_1$ decreases as both the charge $Q$ and Lorentz-violation parameter $l$ increase. This result suggests that additional photon rings contract toward the inner region of the ATSW shadow with increasing values of $Q$ and $l$. Additionally, it is noted that $n_1$, $n_2$, and $n_3$ are more significantly influenced by the Lorentz-violation parameter $l$ than by the charge $Q$.

\subsection{Observed intensity and transfer function}

\begin{figure*}[htbp]
	\centering
	\subfigure[$Q=0.1$.]{
		\begin{minipage}[t]{0.28\linewidth}
			\includegraphics[width=2.1in]{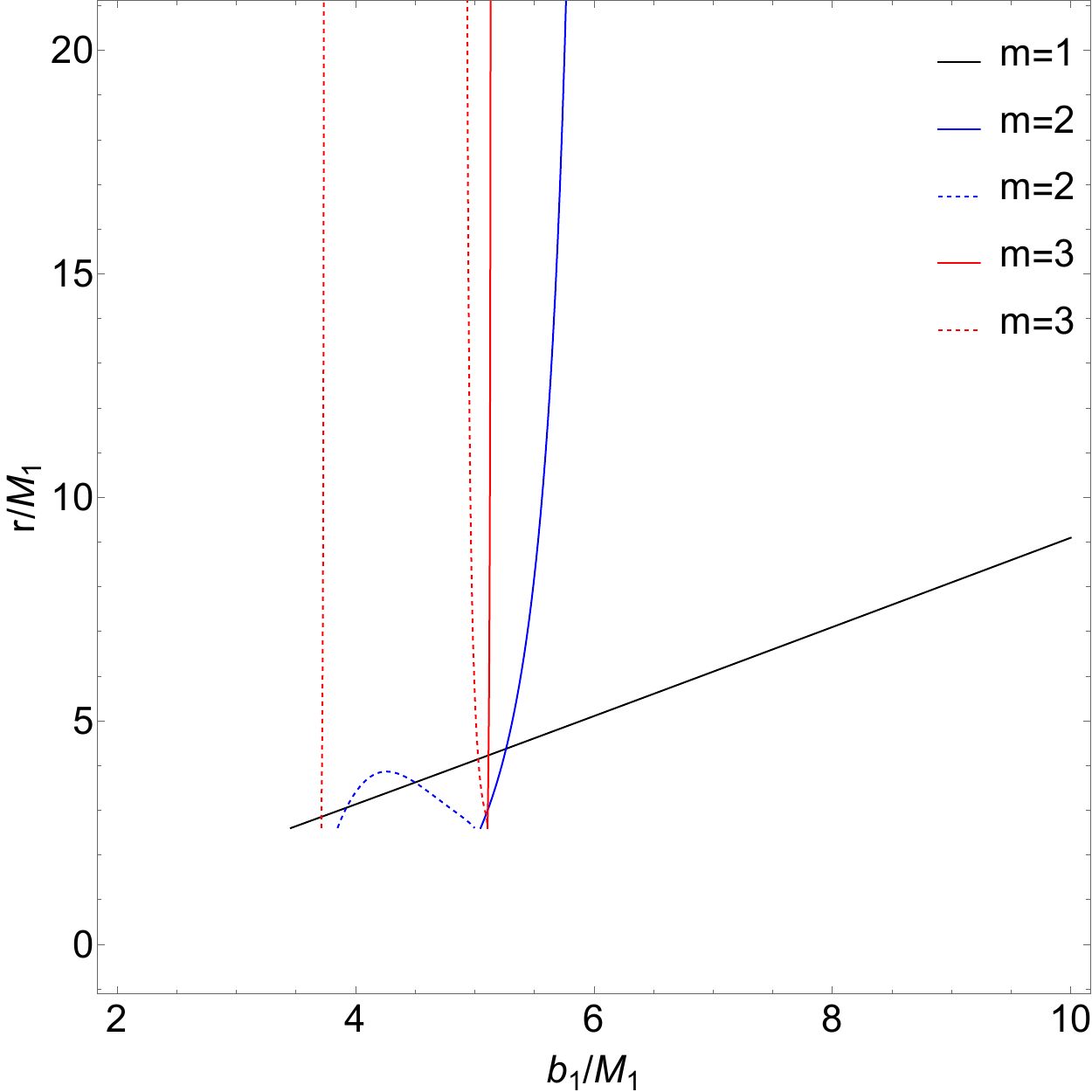}\\
		\end{minipage}
	}
	\subfigure[$Q=0.3$.]{
		\begin{minipage}[t]{0.28\linewidth}
			\includegraphics[width=2.1in]{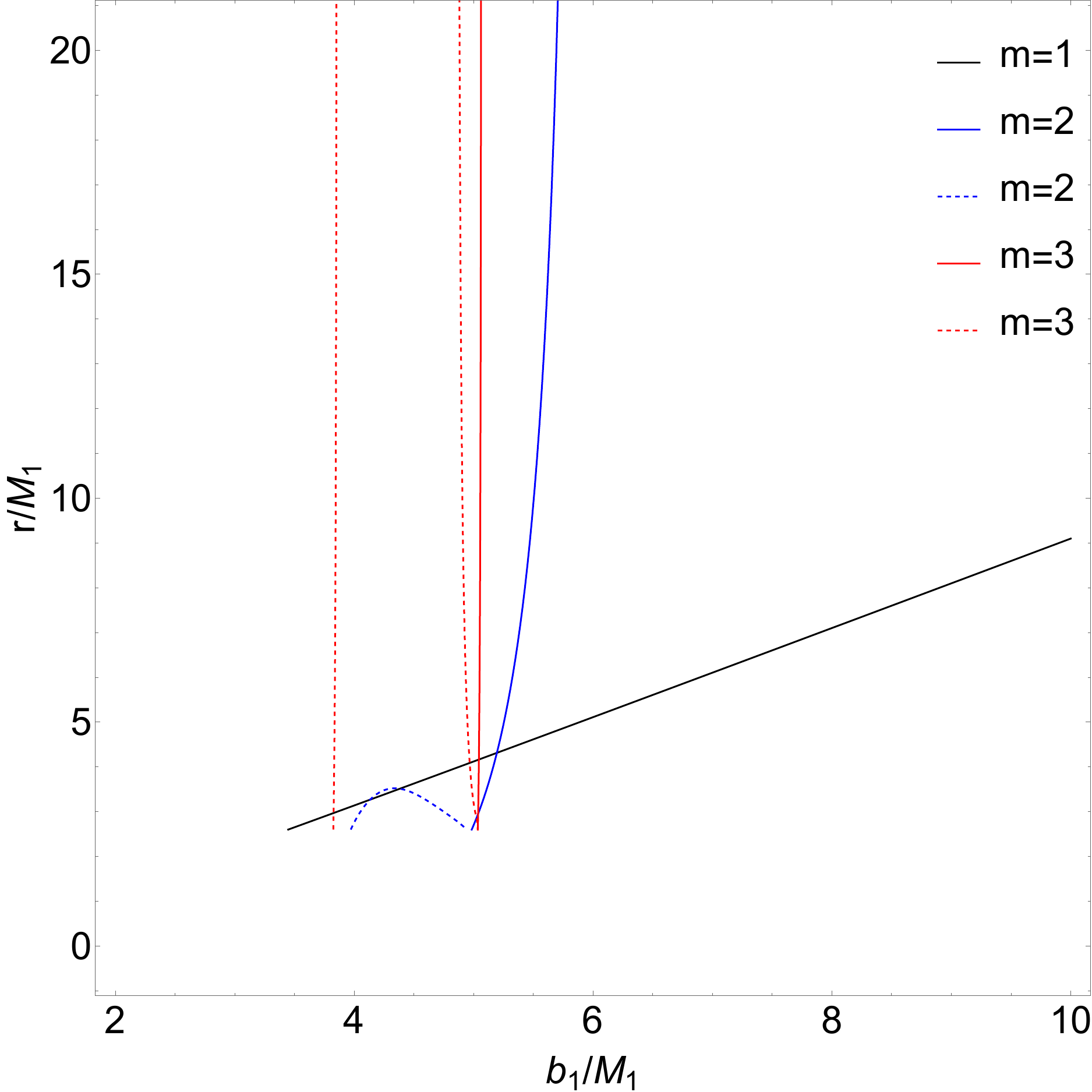}\\
		\end{minipage}
	}
    \subfigure[$Q=0.5$.]{
		\begin{minipage}[t]{0.28\linewidth}
			\includegraphics[width=2.1in]{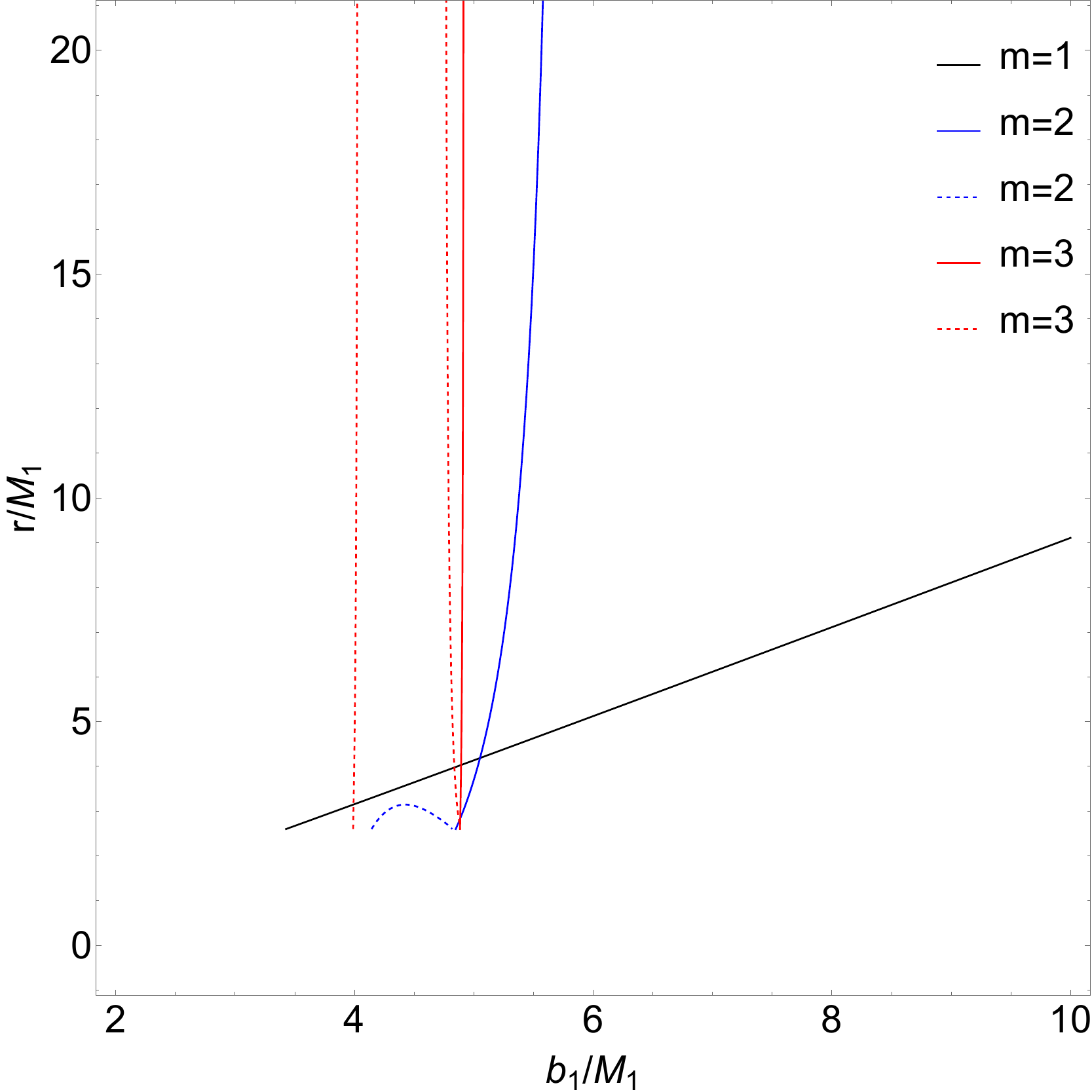}\\
		\end{minipage}
	}
	\caption{(color online) Transfer functions of the ATSW in KR field ($l=0.01$) for different values of $Q$. The black, blue, and red curves in the figure represent the first, second, and third transfer functions, respectively. The parameters are set as $R = 2.6$, $M_1 = 1$, and $M_2 = 1.2$.}
	\label{A7}
\end{figure*}

\begin{figure*}[htbp]
	\centering
	\subfigure[$l=0.01$.]{
		\begin{minipage}[t]{0.28\linewidth}
			\includegraphics[width=2.1in]{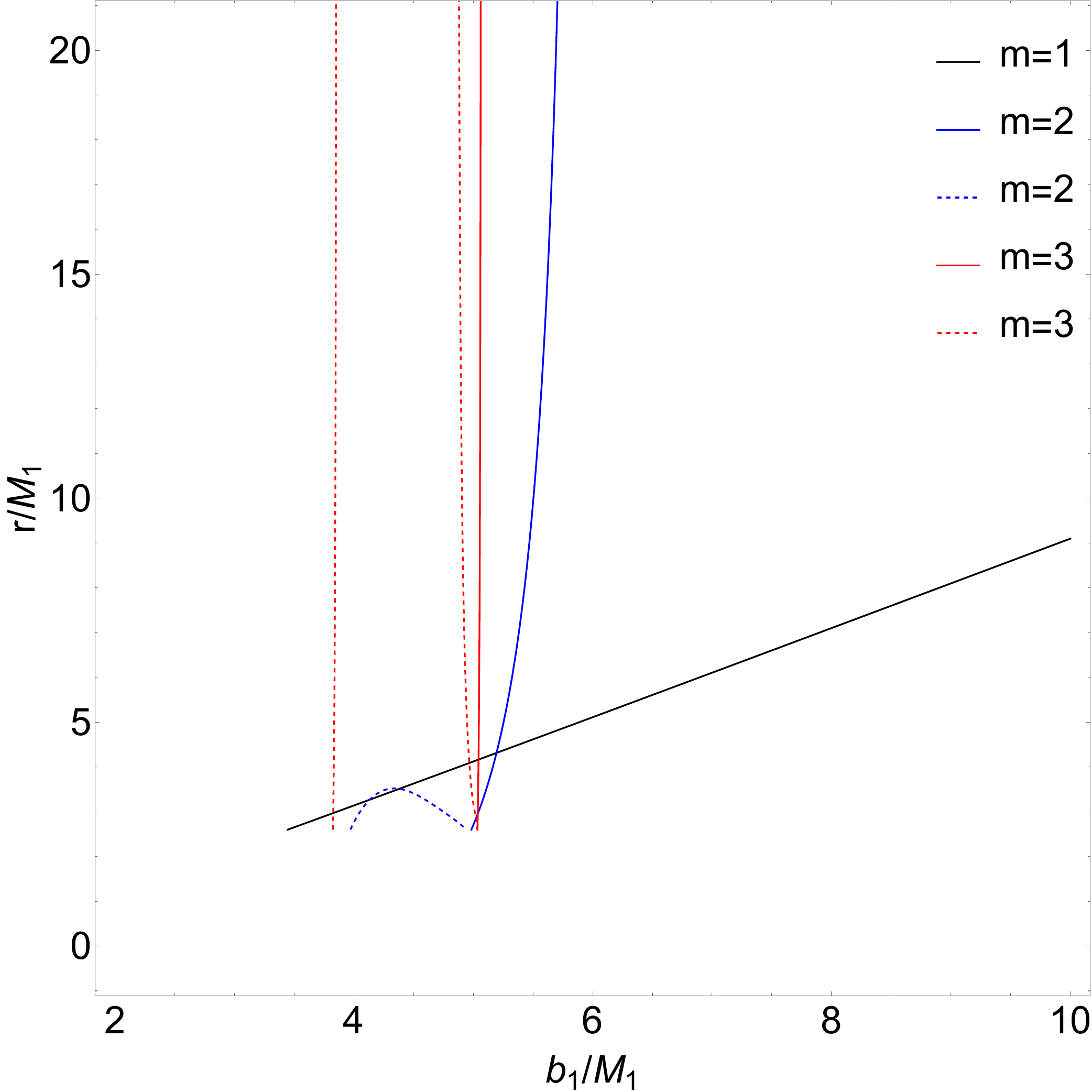}\\
		\end{minipage}
	}
	\subfigure[$l=0.05$.]{
		\begin{minipage}[t]{0.28\linewidth}
			\includegraphics[width=2.1in]{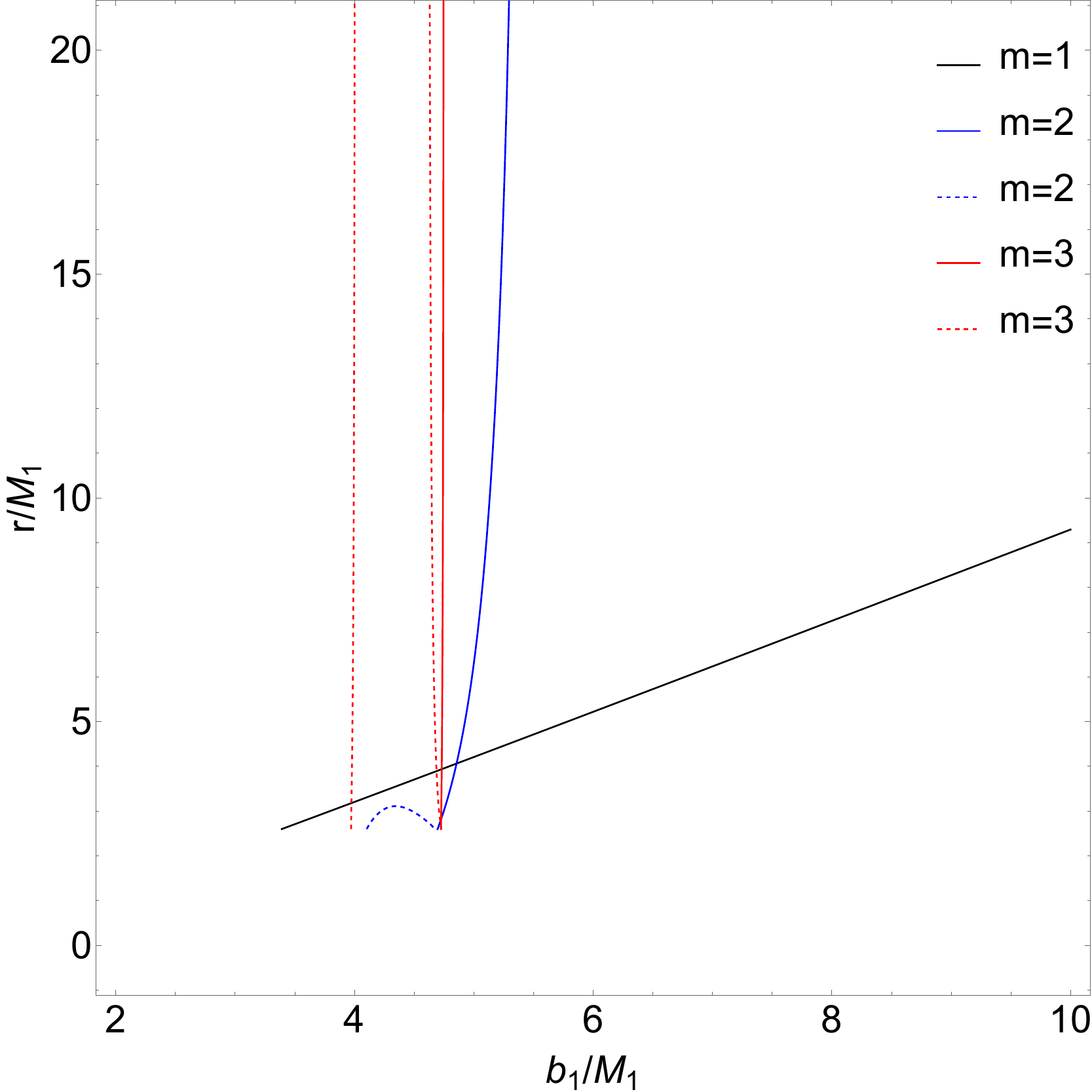}\\
		\end{minipage}
	}
    \subfigure[$l=0.1$.]{
		\begin{minipage}[t]{0.28\linewidth}
			\includegraphics[width=2.1in]{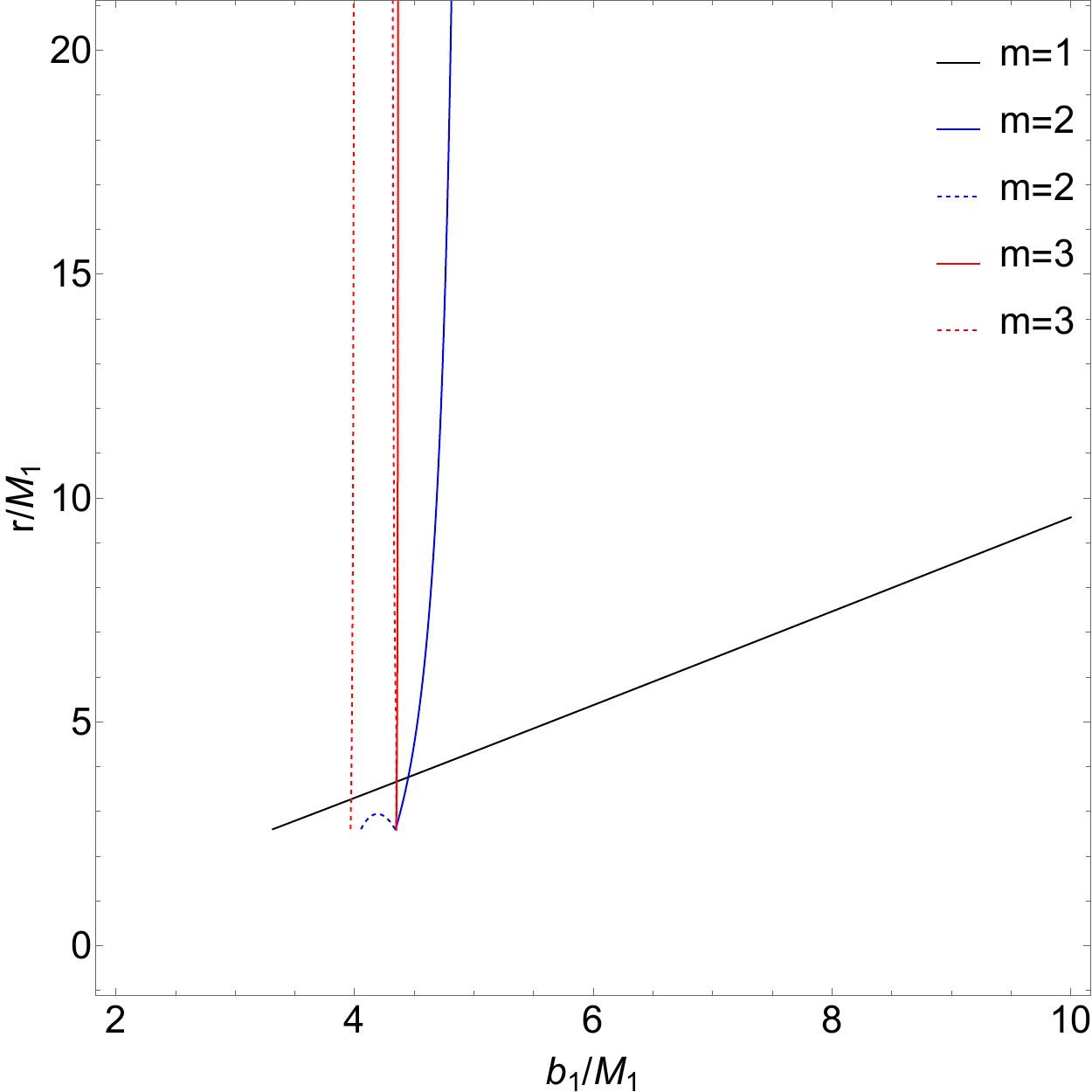}\\
		\end{minipage}
	}
	\caption{(color online) Transfer functions of the ATSW in KR field ($Q=0.3$) for different values of $l$. The black, blue, and red curves in the figure represent the first, second, and third transfer functions, respectively. The parameters are set as $R = 2.6$, $M_1 = 1$, and $M_2 = 1.2$.}
	\label{A8}
\end{figure*}

To better understand the observable appearance of the ATSW, we model the radiation source as an optically and geometrically thin accretion disk located in spacetime $M_1$ of the ATSW. As previously stated, a static observer is positioned at infinity along the northern polar direction. Here, we maintain the assumption that the static observer in spacetime $M_1$ is located in the same polar orientation, while the thin accretion disk resides in the equatorial plane. Owing to the symmetry of spacetime and the isotropic emission from the accretion disk, the radiation intensity depends solely on the radial coordinate and is denoted as $I_\nu^{\mathrm{em}}(r)$, where $v$ represents the emission frequency in the static frame. Thus, we obtain
\begin{equation}\label{eq21}
\frac{I_{\nu^{\prime}}^{\mathrm{obs}}}{I_\nu^{\mathrm{em}}}=\frac{\nu^{\prime 3}}{\nu^3}.
\end{equation}
Here, $\nu^{\prime}$ refers to the redshifted frequency. Under the assumption of a static emitter, this is denoted as $\nu^{\prime}=\sqrt{f} \nu$, which consequently yields the observed intensity
\begin{equation}\label{eq22}
\frac{I_{\nu^{\prime}}^{\mathrm{obs}}}{I_\nu^{\mathrm{em}}(r)}=f^{3 / 2}(r) .
\end{equation}
The frequency shift factor thus accounts solely for gravitational redshift in this context. Based on the Eq.~(\ref{eq22}), we integrated the observed intensity to obtain the total observed intensity at the intersection points
\begin{equation}\label{eq23}
I^{\mathrm{obs}}=\int I_{\nu^{\prime}}^{\mathrm{obs}} \mathrm{~d} \nu^{\prime}=\int f^2 I_\nu^{\mathrm{em}} \mathrm{~d} v=f^2(r) I^{\mathrm{em}}(r).
\end{equation}
Here, $I^{\mathrm{em}}(r)=\int I_\nu^{\mathrm{em}} \mathrm{d} \nu$ represents the total intensity radiated from the accretion disk. Consequently, we obtain the total observed intensity from all intersection points as a function of the impact parameter $b_1$ 
\begin{equation}\label{eq24}
I_{\mathrm{obs}}(b_1)=\left.\sum_m I^{\mathrm{em}}(r) f^2(r)\right|_{r=r_m\left(b_1\right)}.
\end{equation}
With $r_m\left(b_1\right)$ represents the transfer function, and $b_1$ denotes the impact parameter. The radial coordinate of the intersection point between the thin accretion disk and the photon trajectory is determined by the transfer function $r_m\left(b_1\right)$. When $m=1$, $m=2$, and $m=3$, they correspond to the direct emission, lensing ring, and photon ring of the thin accretion disk, respectively. The relationship between the transfer function and the impact parameter is illustrated in Figs.~\ref{A7}~ and~\ref{A8}~, where we have considered the effects of varying charge and Lorentz -violation parameters.

Figs.~\ref{A7}~ and ~\ref{A8}~ show the relationship between the transfer function and the impact parameter $b_1$ under the influence of the charge $Q$ and the Lorentz-violation parameter $l$. The selected values for $Q$ are $0.1$, $0.3$, and $0.5$, while for $l$ they are $0.01$, $0.05$, and $0.1$. From Figs.~\ref{A7}~ and ~\ref{A8}~, it can be observed that the first transfer function ($m=1$), represented by the black solid line, corresponds to the direct image of the accretion disk. Due to its relatively small slope-where $\mathrm{d} r (b_1) / \mathrm{d} b_1$ is defined as the demagnification factor, the small demagnification leads to the observed intensity being dominated by the direct image. The second transfer function ($m=2$), shown as the blue solid line, exhibits a larger demagnification factor, forming a demagnified lensing ring from the back side of the thin accretion disk and contributing only marginally to the observed intensity. The third transfer function ($m=3$), represented by the red solid line, displays the largest demagnification factor, producing a significantly demagnified photon ring from the front side of the thin accretion disk.

As shown in Figs.~\ref{A7}~ and ~\ref{A8}~, a comparison of the transfer functions between the BH and the ATSW reveals that the ATSW exhibits two additional transfer functions near the critical curves $Zb_{c_2}$ and $b_{c_1}$, represented by the blue and red dashed lines in the figures. The slopes of the new second and third transfer functions are approximately the same as those of the conventional second and third transfer functions, respectively. In summary, the additional transfer functions in the ATSW produce extra light rings, a feature that distinguishes it from the BH. Furthermore, we observe that as the charge $Q$ and the Lorentz-violation parameter $l$ increase, the positions of all transfer functions shift inward, with the second and third transfer functions undergoing more significant changes. It is also noted that the Lorentz-violation parameter $l$ has a more pronounced influence on the transfer functions than the charge $Q$.

\subsection{Optical Appearance of Thin Accretion Disks under Two Emission Models in an Asymmetric Thin-Shell Wormhole}

To better understand the optical appearance of the ATSW in a KR field, this section investigates its observable features through two distinct emission models of a thin accretion disk. The optical appearance of the black hole observed by a static observer at infinity is also determined by the emission model of the thin accretion disk\cite{Wang:2025opr}.\\

In emission model $A$, the radiation profile is defined by the following expression
\begin{equation}\label{eq25}
I_A^{\mathrm{em}}(r)= \begin{cases}0 & r<r_{i s c o}, \\ \left(\frac{1}{r-\left(r_{i s c o}-1\right)}\right)^2 & r \geq r_{i s c o}.\end{cases}
\end{equation}
Here, $r_{i s c o}$ denotes the innermost stable circular orbit (ISCO) of massive particles, defined as
\begin{equation}\label{eq26}
r_{\text {isco }}=\frac{3 f\left(r_{\text {isco }}\right) f^{\prime}\left(r_{\text {isco }}\right)}{2 f^{\prime}\left(r_{\text {isco }}\right)^2-f\left(r_{\text {isco }}\right) f^{\prime \prime}\left(r_{\text {isco }}\right)} .
\end{equation}

\begin{figure*}[htbp]
	\centering
	\subfigure[Emission model $A$.]{
		\begin{minipage}[t]{0.33\linewidth}
			\includegraphics[width=2.4in]{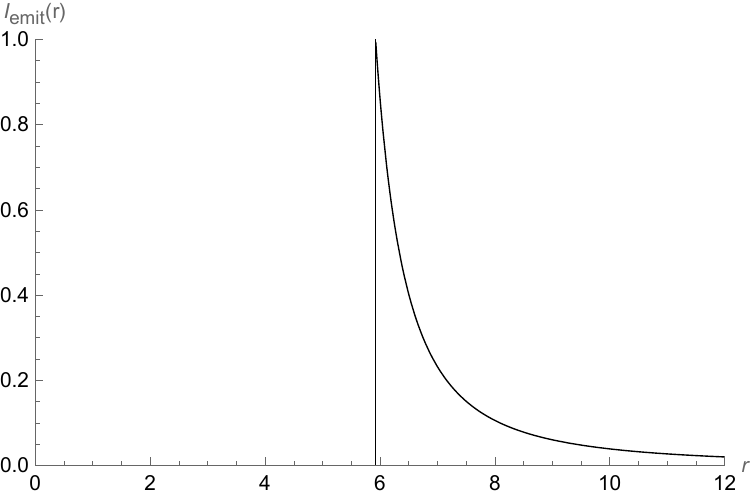}\\
		\end{minipage}
	}
	\subfigure[Emission model $B$.]{
		\begin{minipage}[t]{0.33\linewidth}
			\includegraphics[width=2.4in]{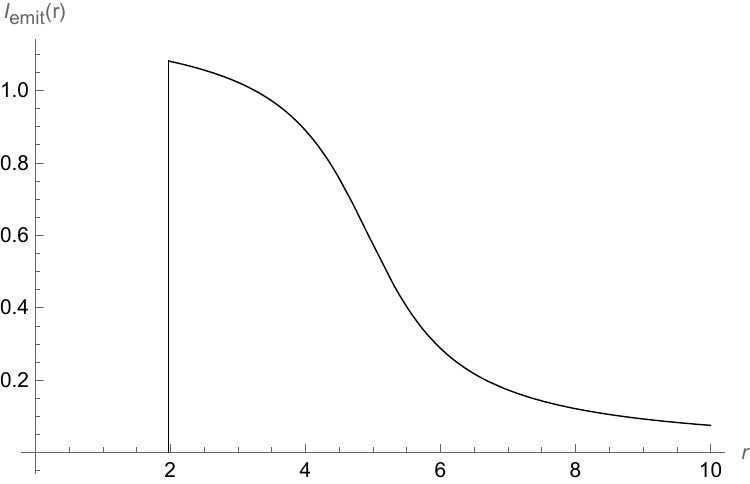}\\
		\end{minipage}
	}
	\caption{(color online) Function plots for the two emission models $A$ and $B$ of a thin accretion disk.}
	\label{A9}
\end{figure*}
\begin{figure*}[htbp]
	\centering
	\subfigure[]{
		\begin{minipage}[t]{0.28\linewidth}
			\includegraphics[width=2.1in]{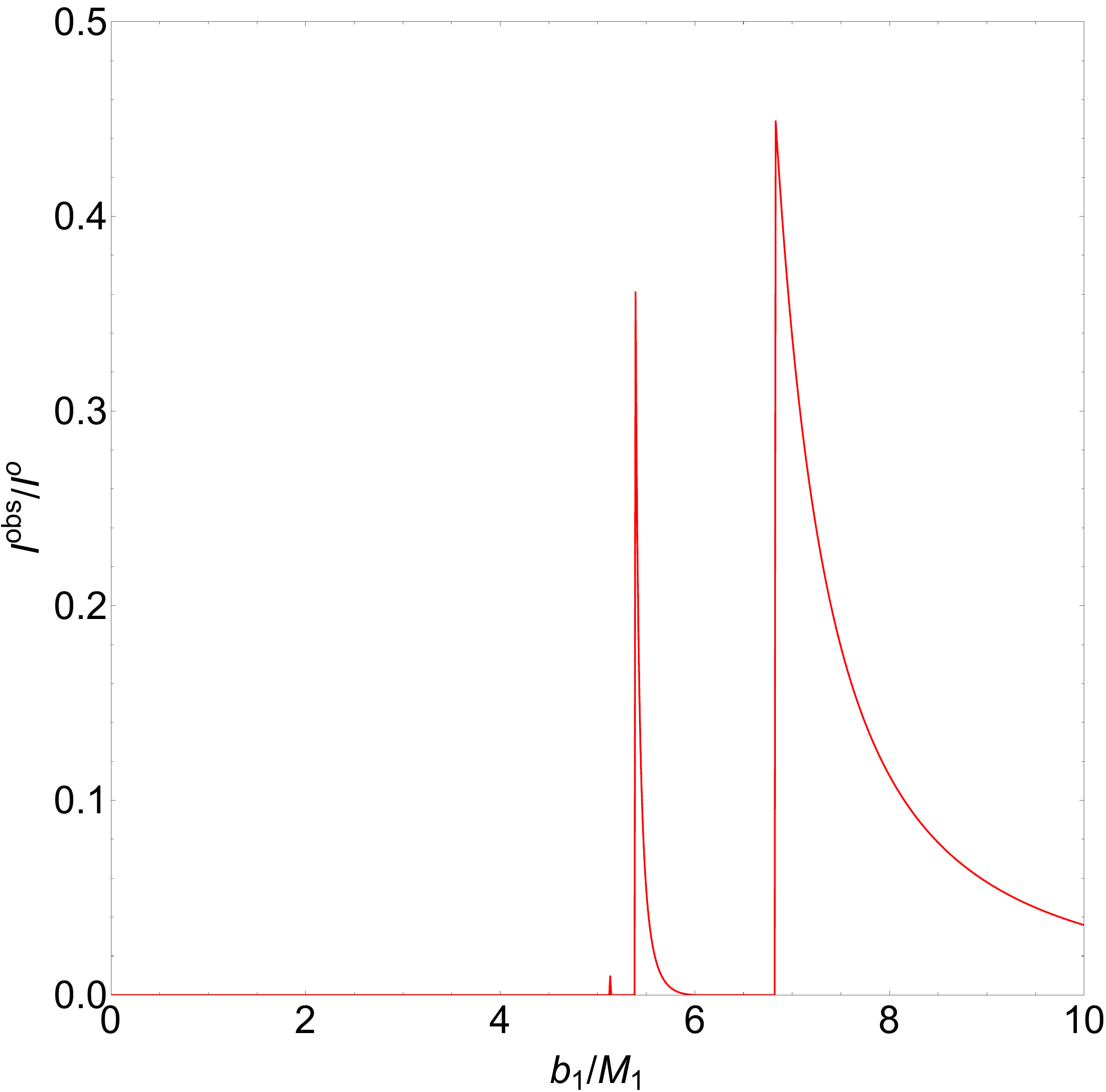}\\
		\end{minipage}
	}
	\subfigure[]{
		\begin{minipage}[t]{0.28\linewidth}
			\includegraphics[width=2.1in]{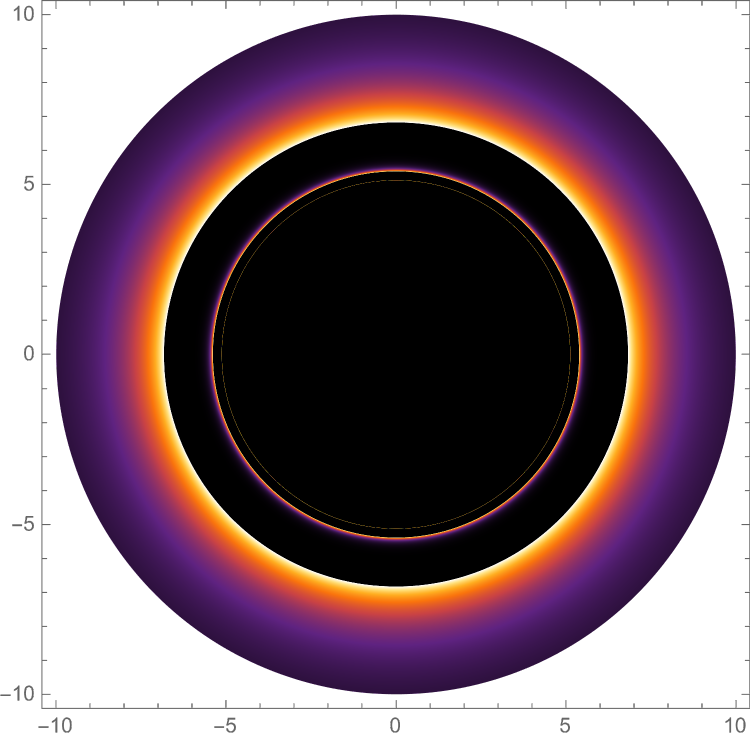}\\
		\end{minipage}
	}
    \subfigure[]{
		\begin{minipage}[t]{0.28\linewidth}
			\includegraphics[width=2.1in]{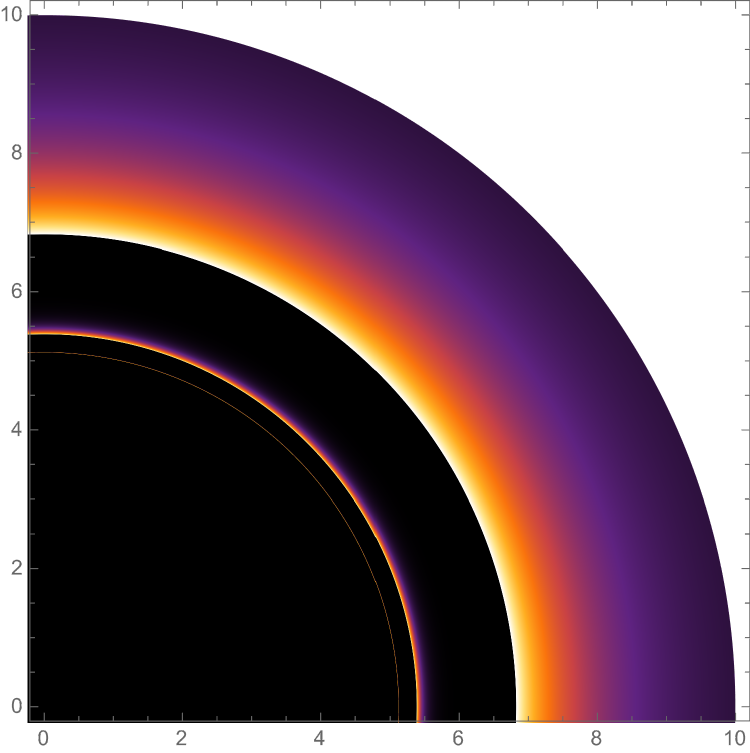}\\
		\end{minipage}
	}
    \subfigure[]{
		\begin{minipage}[t]{0.28\linewidth}
			\includegraphics[width=2.1in]{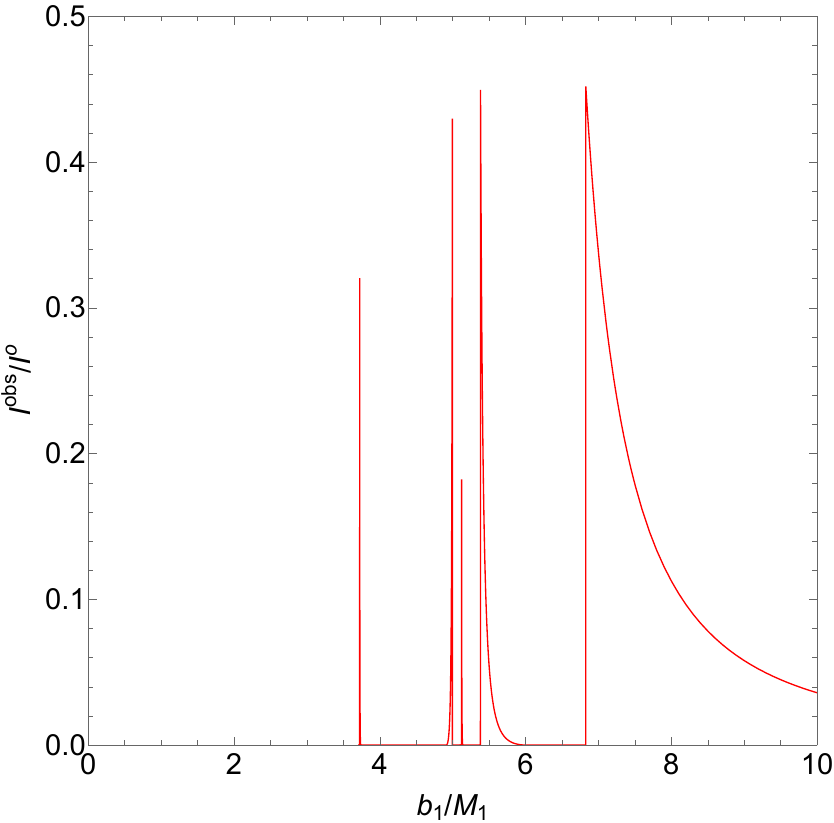}\\
		\end{minipage}
	}
	\subfigure[]{
		\begin{minipage}[t]{0.28\linewidth}
			\includegraphics[width=2.1in]{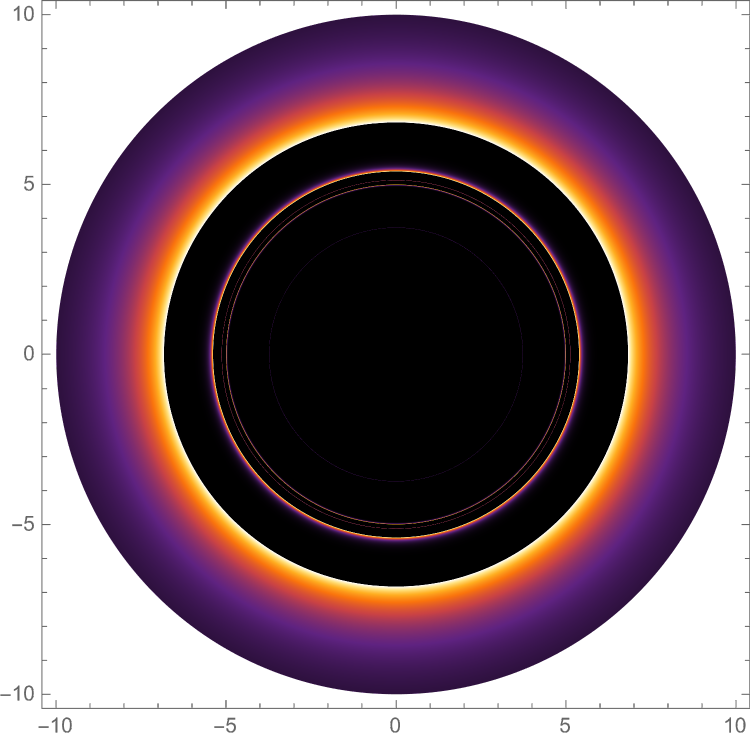}\\
		\end{minipage}
	}
    \subfigure[]{
		\begin{minipage}[t]{0.28\linewidth}
			\includegraphics[width=2.1in]{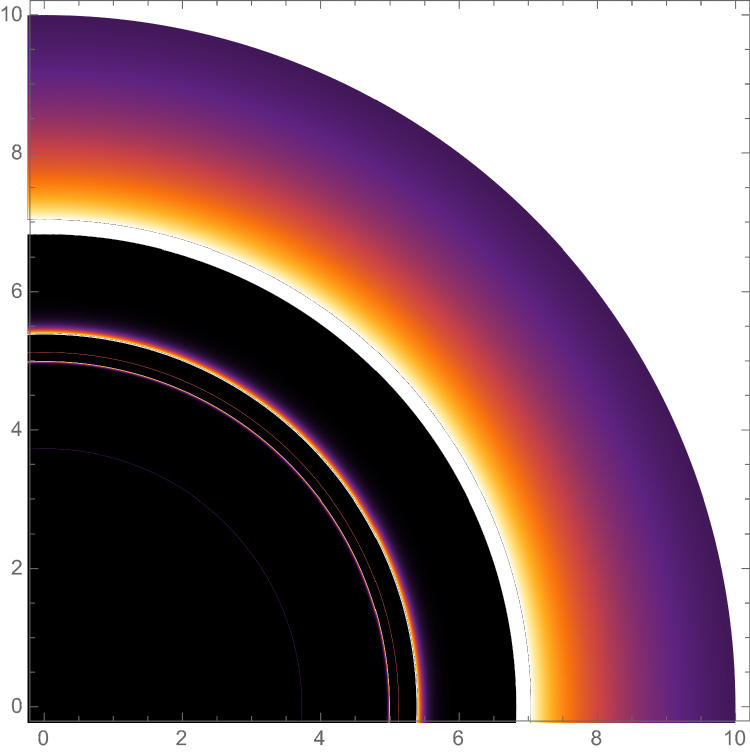}\\
		\end{minipage}
	}
	\caption{(color online) In Model $A$, the observed intensity, density map, and partial density map for a BH with KR structure (top row) and an ATSW (bottom row) are presented. The first, second, and third columns correspond to the observed intensity, density map, and partial density map, respectively. The parameters are set as $Q = 0.1$, $l = 0.01$, $R = 2.6$, $M_1 = 1$, and $M_2 = 1.2$.}
	\label{A10}
\end{figure*}

The radiation flux peaks at $r=r_{i s c o}$ and then decreases with radius, asymptotically approaching zero. For an observer at infinity, the effective radiation originating from within the photon unstable orbit is negligible, as any emission from this region is either highly redshifted or captured by the central object. We have plotted the radiation function, as shown in  Fig.~\ref{A9}~(a). Additionally, we have depicted the optical appearance of black holes with the same mass parameter and radiation model $A$ in the first row of Fig.~\ref{A10}~. For better comparison, we have computed the observed intensity and constructed the corresponding intensity distribution and local intensity maps for the ATSW using Model A through our ray-tracing method. The results are shown in the second row of Fig.~\ref{A10}~. By comparing Figs.~\ref{A10}~(a) and ~\ref{A10}~(b), it is evident that BH and ATSW exhibit significant differences in their optical structures, with distinct spatial separations between the direct radiation ring, lens ring, and photon ring. Observing Fig.~\ref{A10}~(a), we find that the observational intensity of ATSW appears at the critical curve $b_1 \simeq 6.8289 M_1$, with an intensity of $0.4529$, which subsequently decreases to zero. The lens ring is located within a narrow region between the critical curves $b_1 \simeq 5.3774 M_1$ and $b_1 \simeq 5.8891 M_1$. The photon rings appear near the critical curves at $b_1 \simeq 3.7221 M_1$, $b_1 \simeq 4.9676 M_1$, and $b_1 \simeq 5.1213 M_1$. Compared to BH, intriguingly, two new photon rings emerge in the shadow of ATSW, located near the critical curves at $b_1 \simeq 3.7221 M_1$ and $b_1 \simeq 4.9676 M_1$, as shown in Figs.~\ref{A10}~(a) and ~\ref{A10}~(d). By examining Figs.~\ref{A10}~(b) and ~\ref{A10}~(c), we observe that the direct emission ring is near the edge of the black disk, while the lens ring is positioned inside the black disk. Furthermore, Figs.~\ref{A10}~(e) and ~\ref{A10}~(f) reveal that the additional photon rings of ATSW are closer to the black disk.\\

\begin{figure*}[htbp]
	\centering
	\subfigure[]{
		\begin{minipage}[t]{0.28\linewidth}
			\includegraphics[width=2.1in]{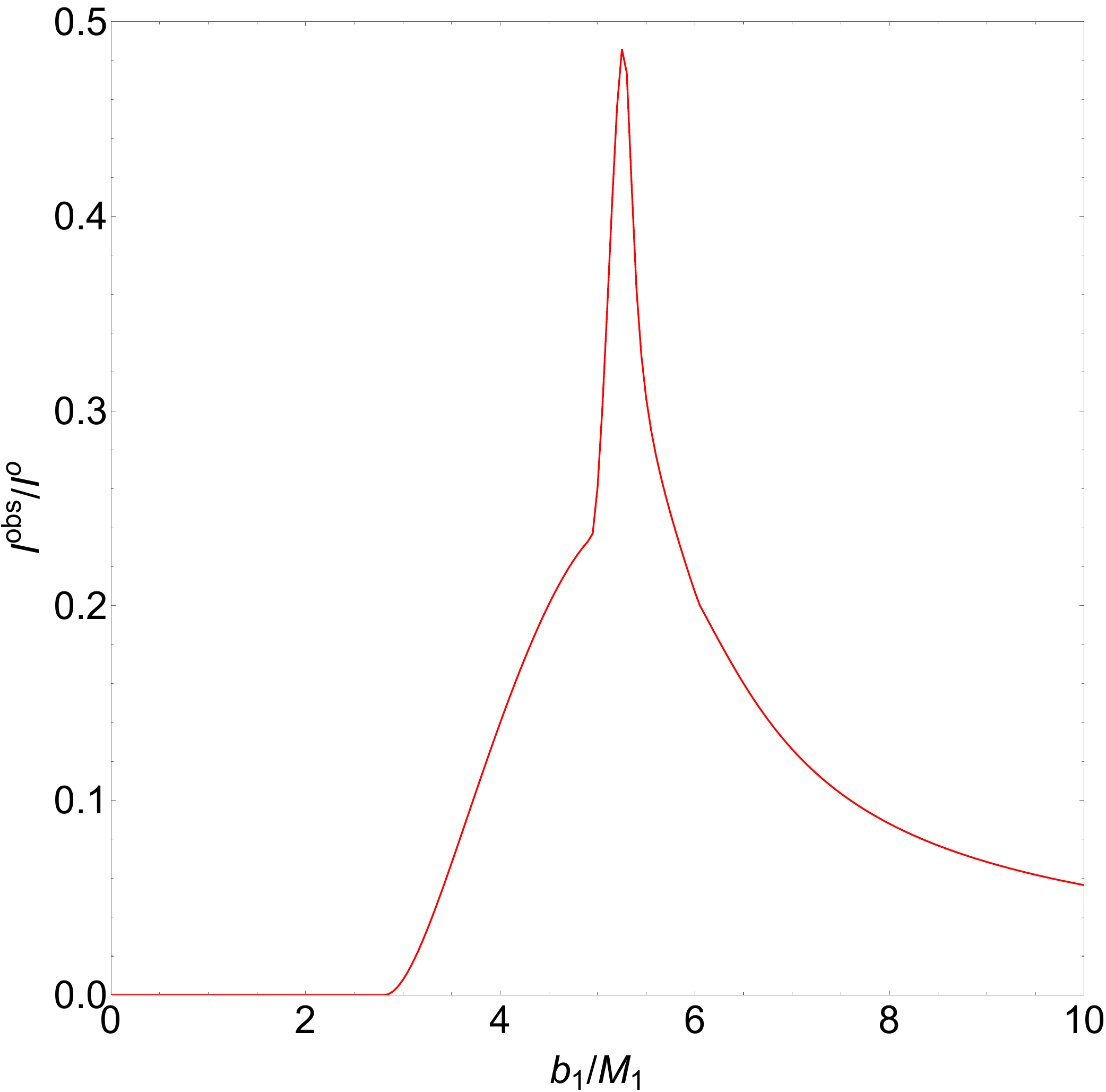}\\
		\end{minipage}
	}
	\subfigure[]{
		\begin{minipage}[t]{0.28\linewidth}
			\includegraphics[width=2.1in]{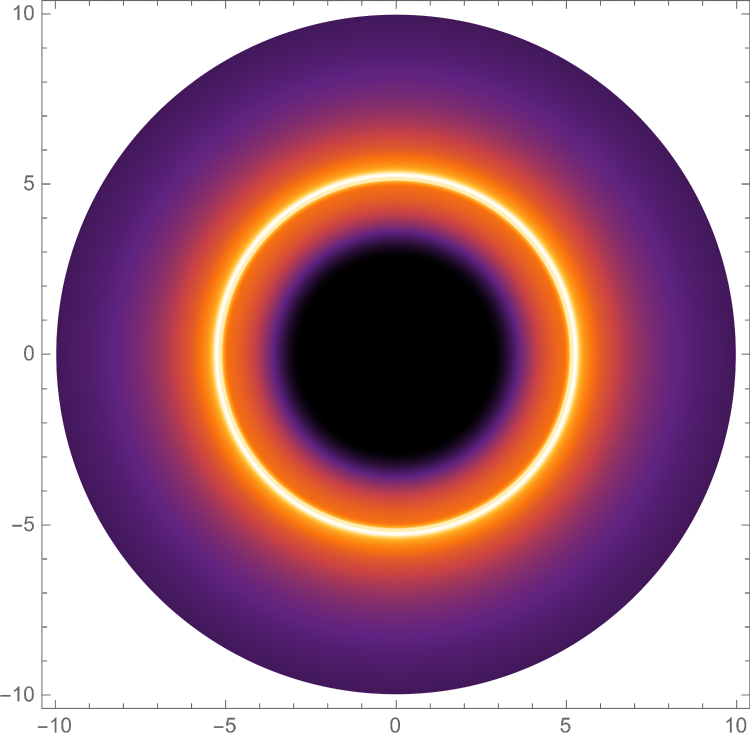}\\
		\end{minipage}
	}
    \subfigure[]{
		\begin{minipage}[t]{0.28\linewidth}
			\includegraphics[width=2.1in]{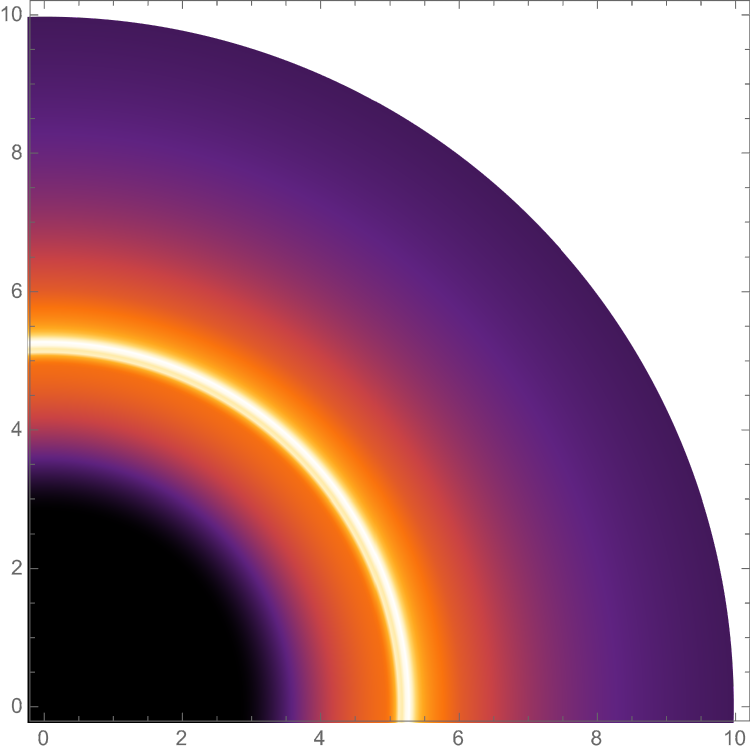}\\
		\end{minipage}
	}
    \subfigure[]{
		\begin{minipage}[t]{0.28\linewidth}
			\includegraphics[width=2.1in]{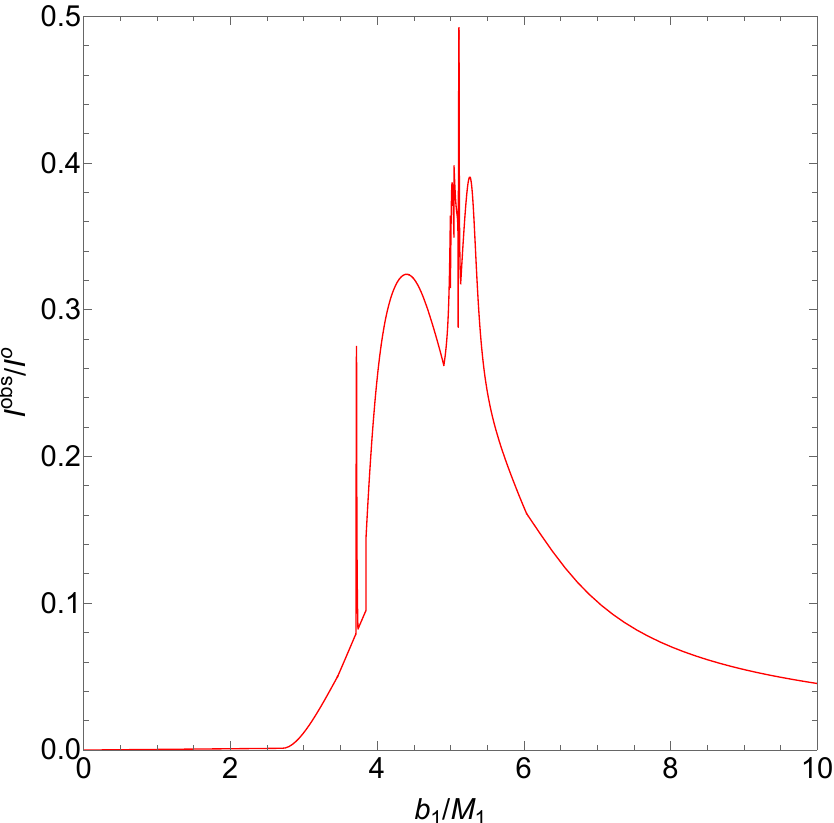}\\
		\end{minipage}
	}
	\subfigure[]{
		\begin{minipage}[t]{0.28\linewidth}
			\includegraphics[width=2.1in]{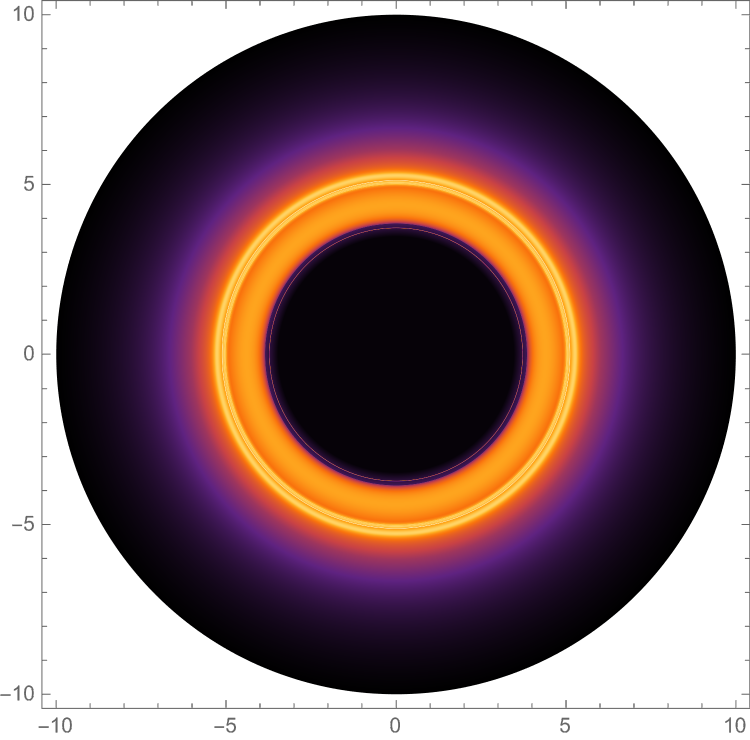}\\
		\end{minipage}
	}
    \subfigure[]{
		\begin{minipage}[t]{0.28\linewidth}
			\includegraphics[width=2.1in]{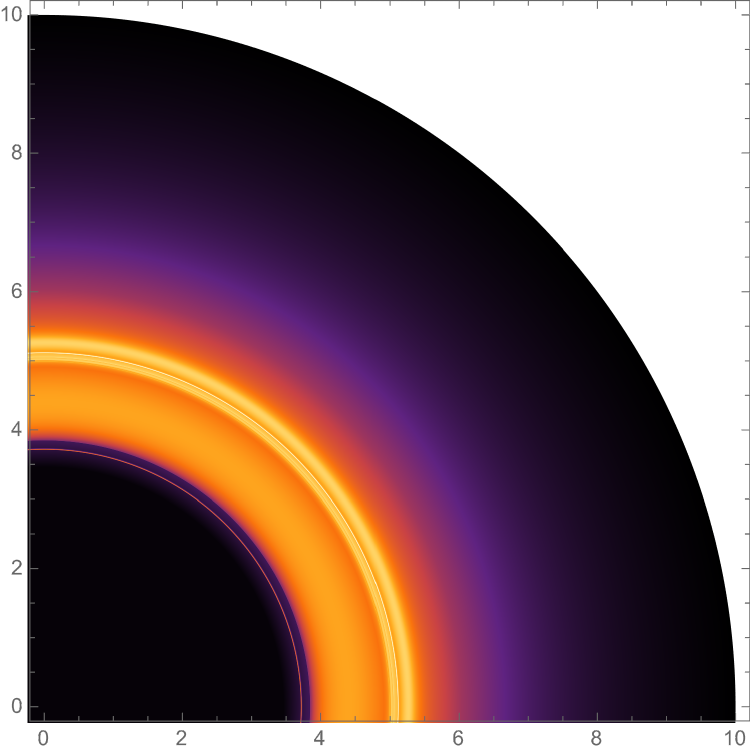}\\
		\end{minipage}
	}
	\caption{(color online) In Model $B$, the observed intensity, density map, and partial density map for a BH with KR structure (top row) and an ATSW (bottom row) are presented. The first, second, and third columns correspond to the observed intensity, density map, and partial density map, respectively. The parameters are set as $Q = 0.1$, $l = 0.01$, $R = 2.6$, $M_1 = 1$, and $M_2 = 1.2$.}
	\label{A11}
\end{figure*}

In emission model $B$, the radiation profile is defined by the following expression
\begin{equation}
I_2^{\mathrm{em}}(r)= \begin{cases}0 & r<r_h, \\ \frac{\frac{\pi}{2}-\tan ^{-1}\left(r-\left(r_{\mathrm{isco}}-1\right)\right)}{\frac{\pi}{2}-\tan ^{-1}\left(r_{p h}\right)} & r \geq r_h.\end{cases}
\end{equation}
In which, $r_h$ represents the event horizon radius. The radiation peak is achieved at $r=r_h$, and then decreases as $r$ increases, approaching zero. The image of radiation model $B$ is shown in Fig.~\ref{A9}~(b), where the decline in radiation value with increasing r is more gradual compared to radiation model $A$. Additionally, we have plotted the optical appearance of black holes with the same mass parameter and radiation model B in the first row of Fig.~\ref{A11}~. For better comparison, we have adopted a numerical method to calculate the observational intensity, intensity distribution, and local intensity distribution of an ATSW using model $B$, as shown in the second row of Figure 10. It can be observed that there is partial overlap in the regions of the direct radiation ring, lens ring, and photon ring in the ATSW. The direct ring in the ATSW starts radiating at $b_1 \simeq 2.7449 M_1$. Interestingly, the lens ring encompasses the photon ring, collectively forming a multi-layered bright ring feature, as illustrated in Figs.~\ref{A11}~(e) and ~\ref{A11}~(f).

\begin{figure*}[htbp]
	\centering
	\subfigure[]{
		\begin{minipage}[t]{0.28\linewidth}
			\includegraphics[width=2.0in]{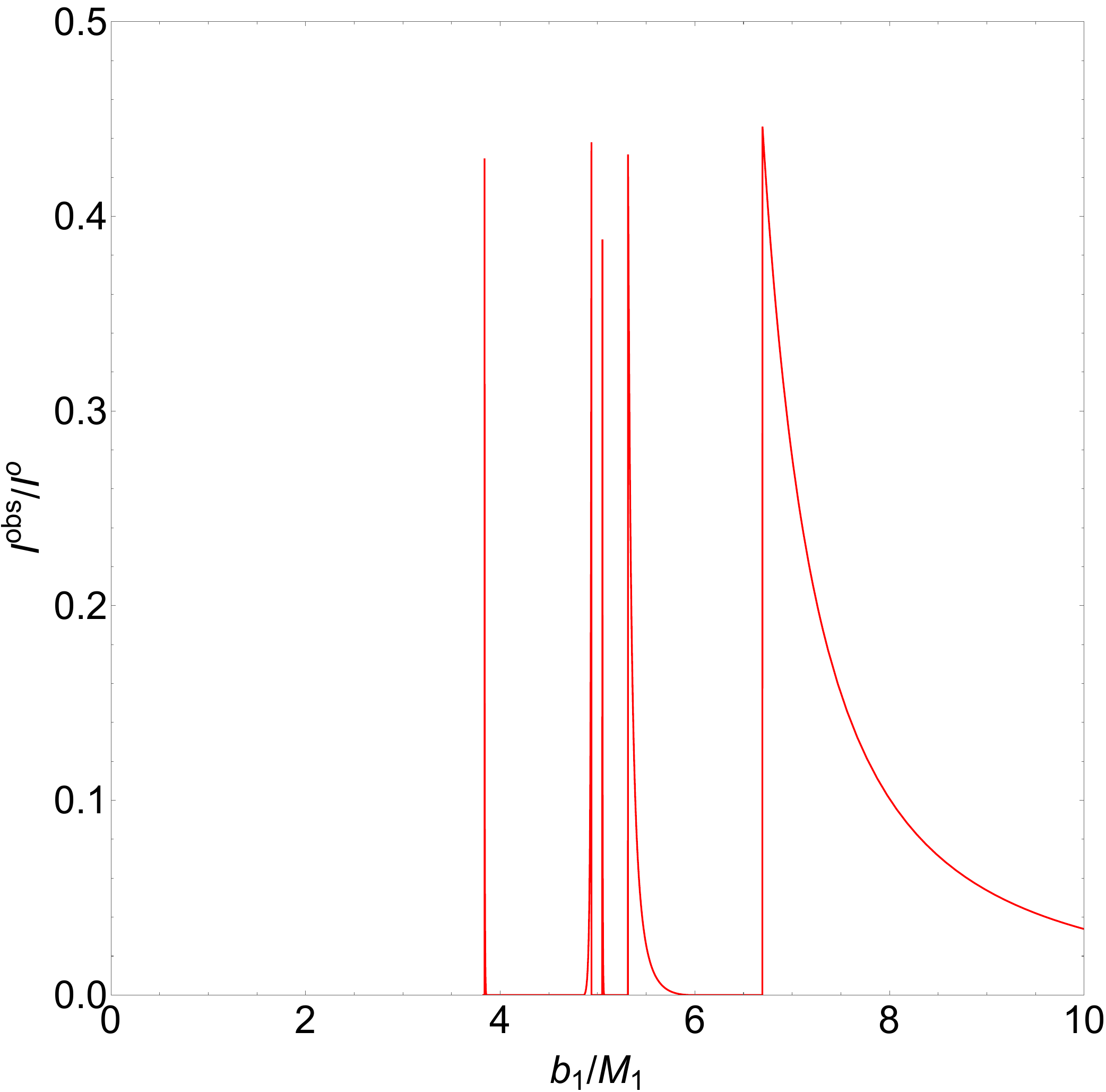}\\
		\end{minipage}
	}
	\subfigure[]{
		\begin{minipage}[t]{0.28\linewidth}
			\includegraphics[width=2.0in]{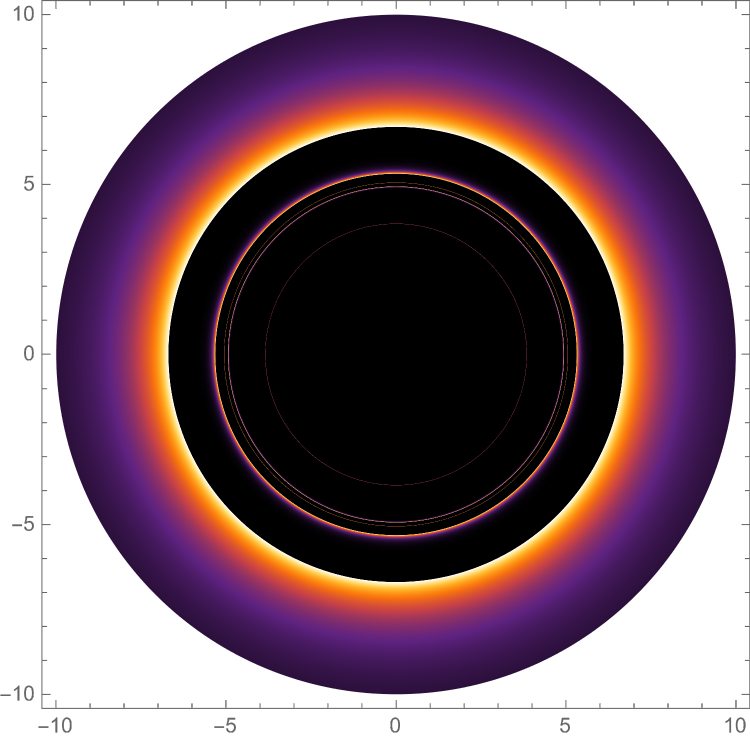}\\
		\end{minipage}
	}
    \subfigure[]{
		\begin{minipage}[t]{0.28\linewidth}
			\includegraphics[width=2.0in]{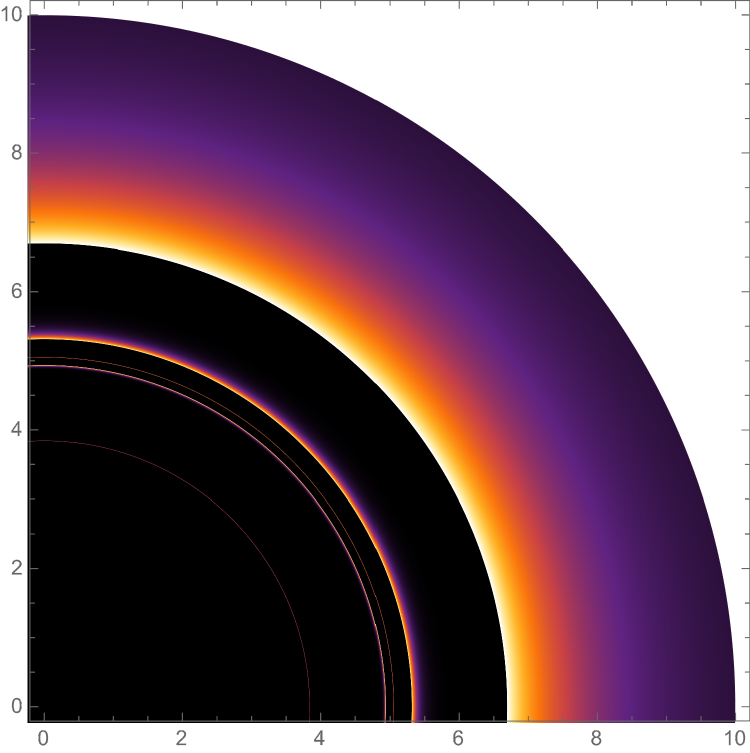}\\
		\end{minipage}
	}
    \subfigure[]{
		\begin{minipage}[t]{0.28\linewidth}
			\includegraphics[width=2.0in]{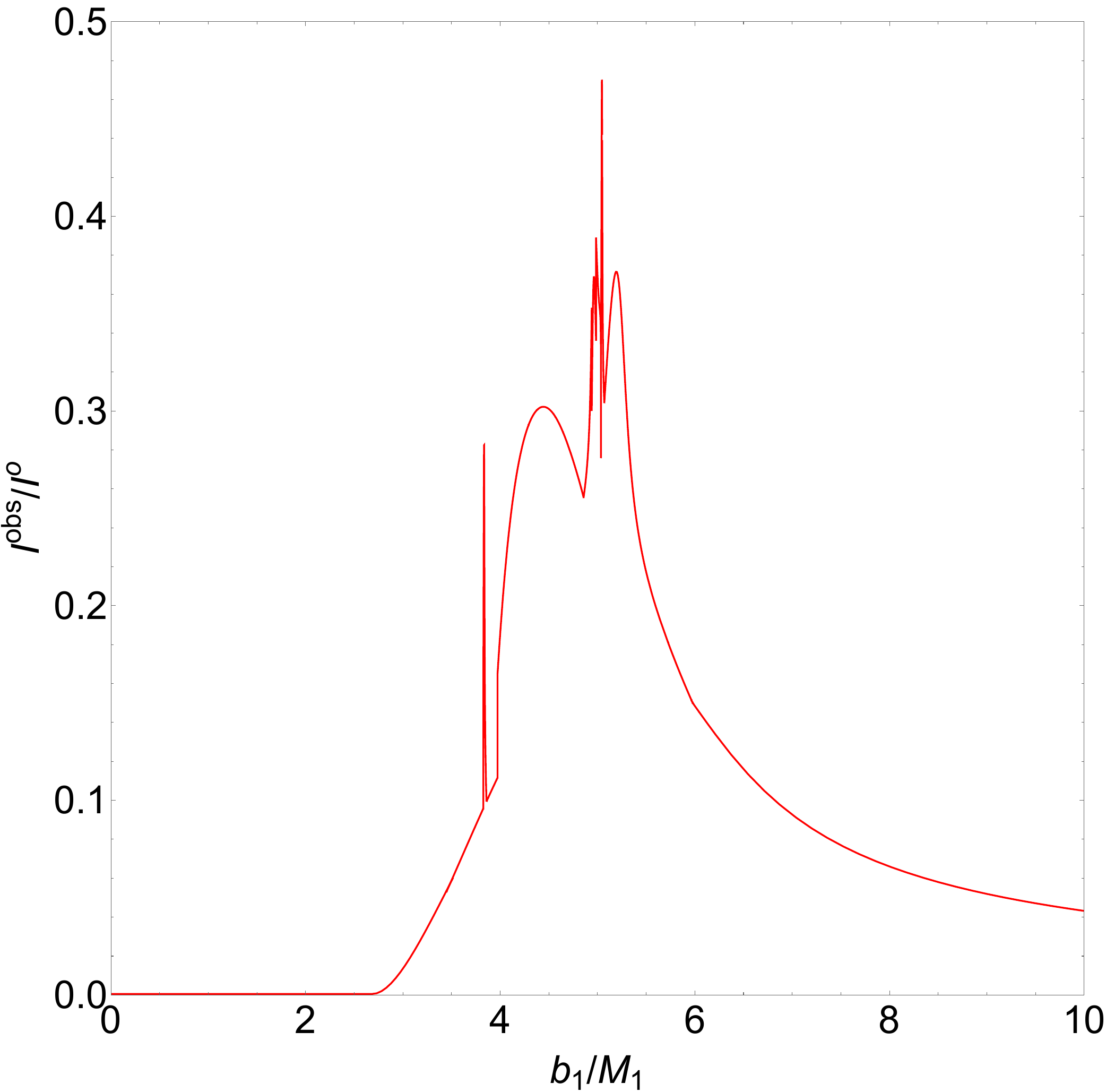}\\
		\end{minipage}
	}
	\subfigure[]{
		\begin{minipage}[t]{0.28\linewidth}
			\includegraphics[width=2.0in]{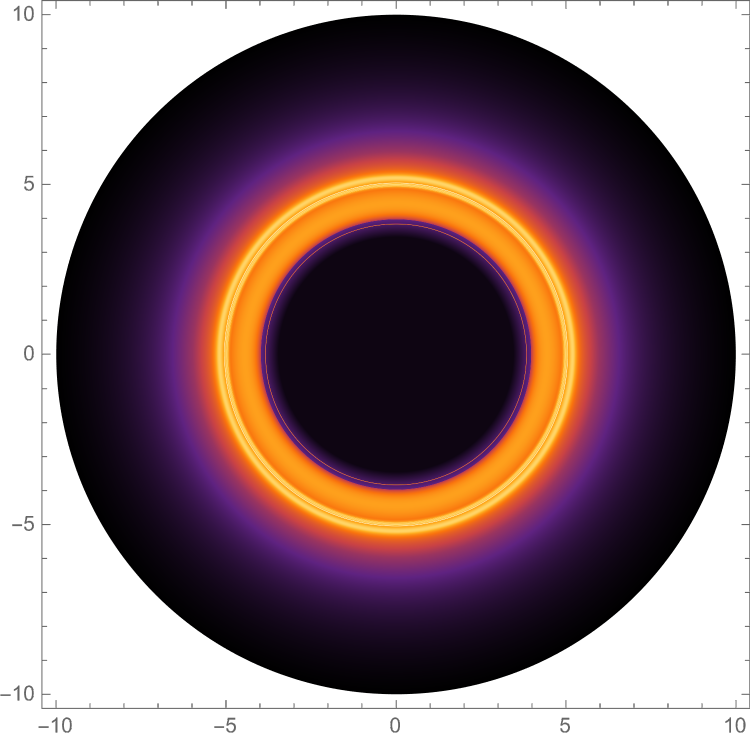}\\
		\end{minipage}
	}
    \subfigure[]{
		\begin{minipage}[t]{0.28\linewidth}
			\includegraphics[width=2.0in]{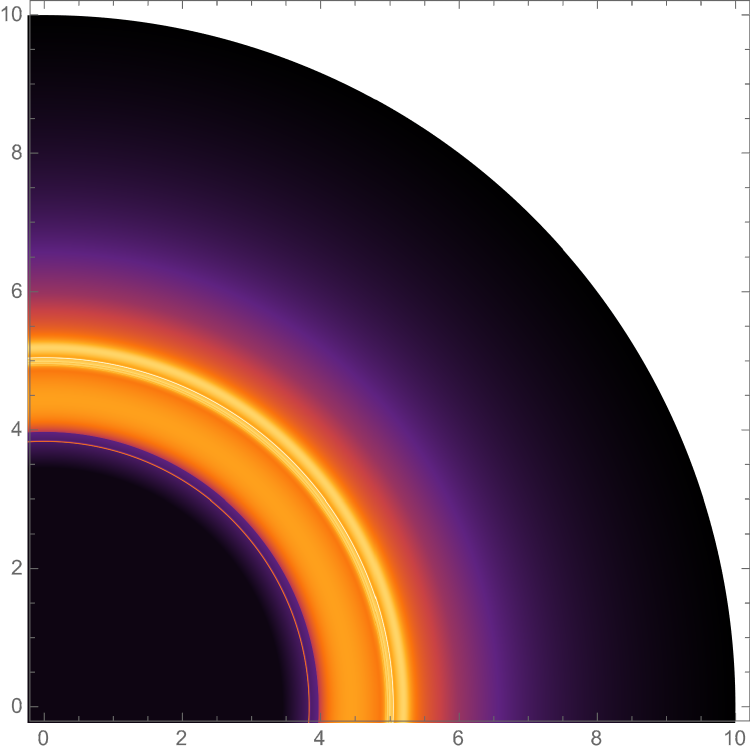}\\
		\end{minipage}
	}
    \subfigure[]{
		\begin{minipage}[t]{0.28\linewidth}
			\includegraphics[width=2.0in]{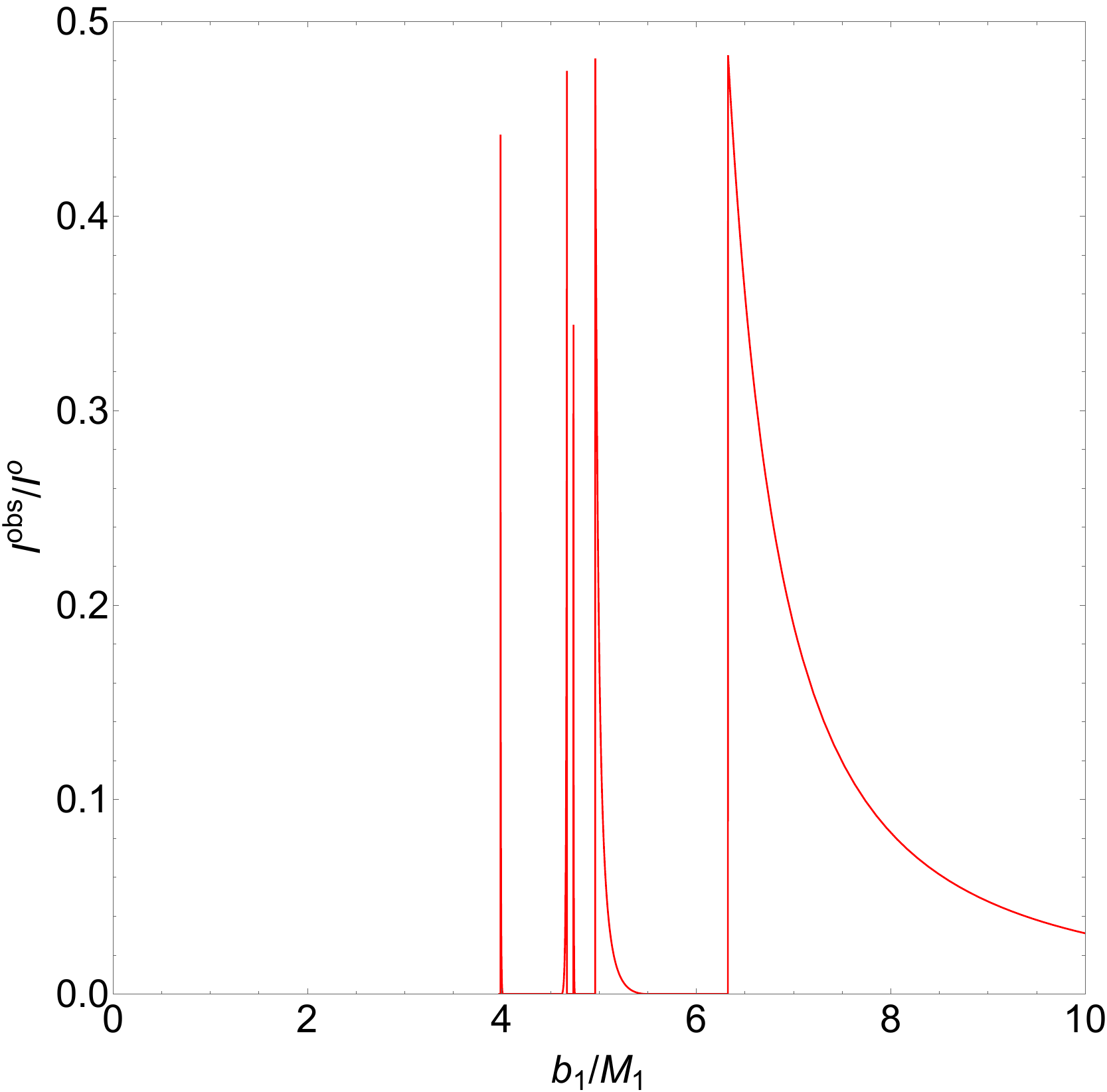}\\
		\end{minipage}
	}
	\subfigure[]{
		\begin{minipage}[t]{0.28\linewidth}
			\includegraphics[width=2.0in]{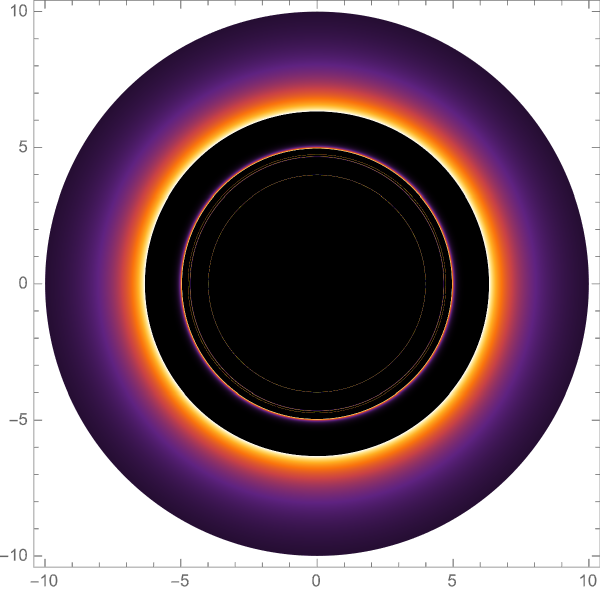}\\
		\end{minipage}
	}
    \subfigure[]{
		\begin{minipage}[t]{0.28\linewidth}
			\includegraphics[width=2.0in]{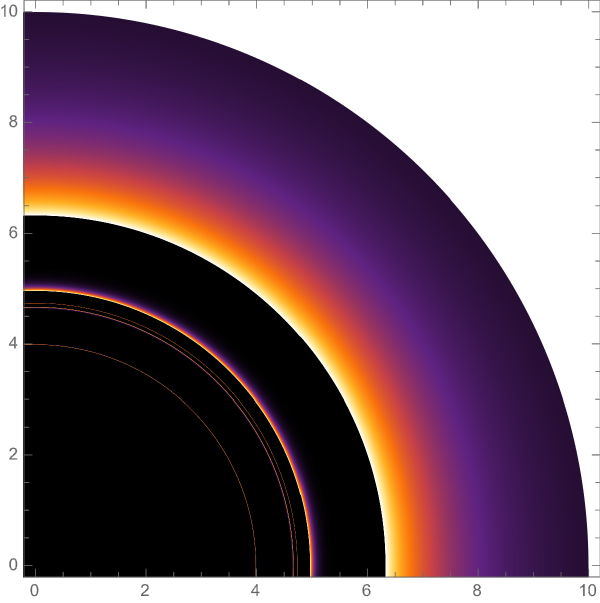}\\
		\end{minipage}
	}
    \subfigure[]{
		\begin{minipage}[t]{0.28\linewidth}
			\includegraphics[width=2.0in]{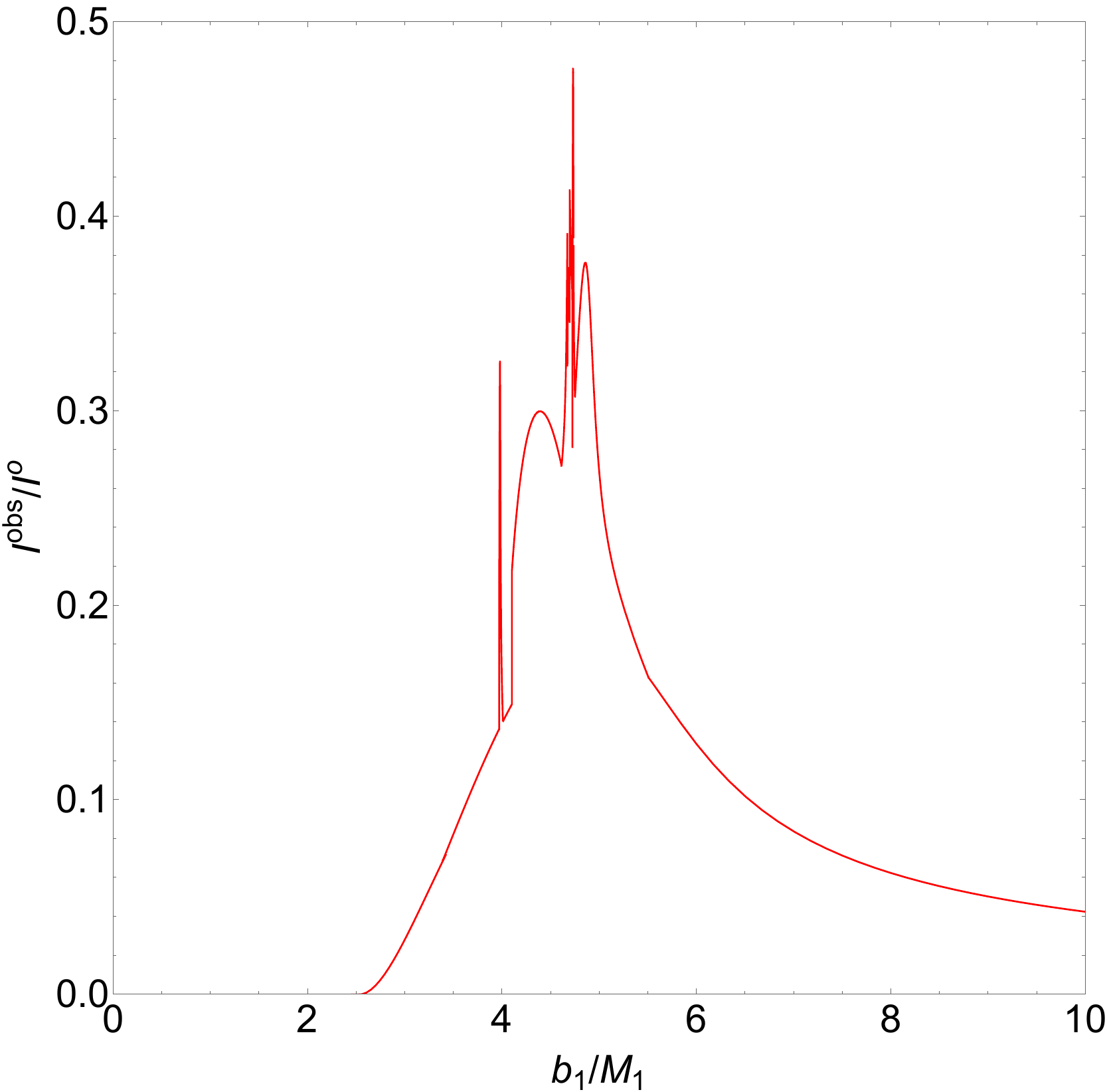}\\
		\end{minipage}
	}
	\subfigure[]{
		\begin{minipage}[t]{0.28\linewidth}
			\includegraphics[width=2.0in]{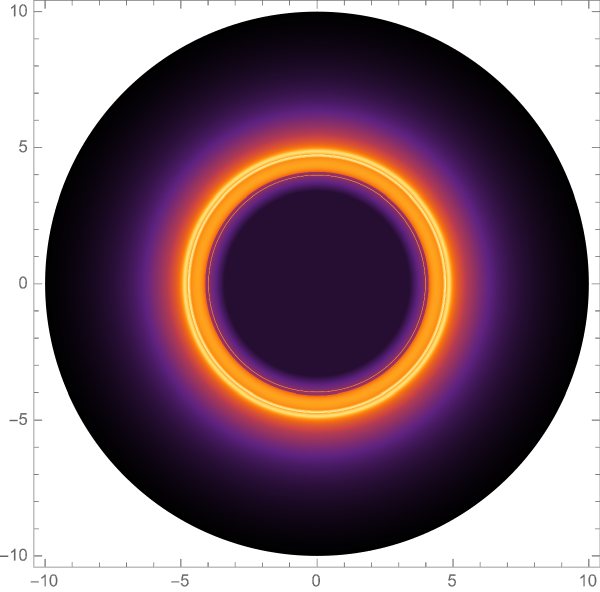}\\
		\end{minipage}
	}
    \subfigure[]{
		\begin{minipage}[t]{0.28\linewidth}
			\includegraphics[width=2.0in]{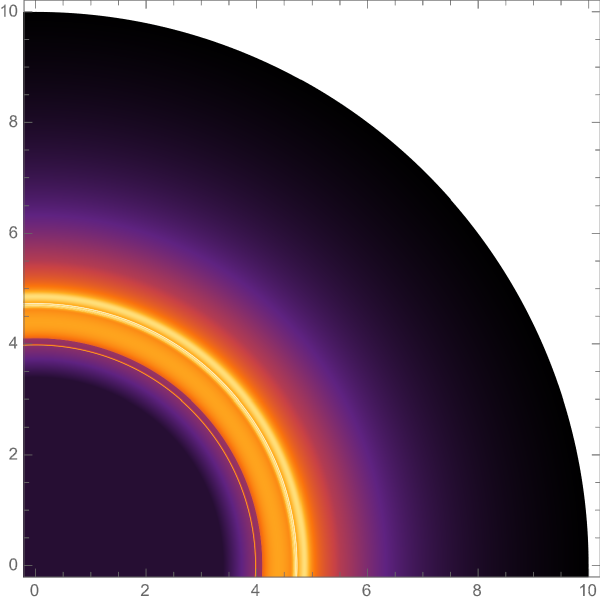}\\
		\end{minipage}
	}
	\caption{(color online) In the ATSW, for the two parameter sets $Q = 0.3$, $l = 0.01$ and $Q = 0.3$, $l = 0.05$, we have plotted the observational intensity (first column), density maps (second column), and local density maps (third column) for two radiation models. The first and second rows of the figure display the observational intensity, density maps, and local density maps for radiation models $A$ and $B$, respectively, at $Q = 0.3$, $l = 0.01$. The third and fourth rows show the same for $Q = 0.3$, $l = 0.05$. We have set $R = 2.6$, $M_1 = 1$, and $M_2 = 1.2$.}
	\label{A12}
\end{figure*}

The difference between radiation models $A$ and $B$ lies in the fact that model $A$ exhibits two additional photon rings near the critical curves at $b_1 \simeq 3.7221 M_1$ and $b_1 \simeq 4.9676 M_1$, whereas model $B$ presents an additional lens ring between the critical curves at $b_1 \simeq 3.7164 M_1$ and $b_1 \simeq 5.1139 M_1$. Consequently, this enhances the observational intensity of ATSW.

We have plotted the observational intensity, density maps, and local density maps for the ATSW in the KR field under two sets of parameters: $Q=0.3$, $l=0.01$ and $Q=0.3$, $l=0.05$, for two radiation models, as shown in Fig.~\ref{A12}~.Observing Figs.~\ref{A10}~.~\ref{A11}~ and ~\ref{A12}~, we find that in radiation model $A$, for the case of $Q=0.3$, $l=0.01$, the two additional photon rings are located at $b_1 \simeq 3.8425 M_1$ and $b_1 \simeq 4.0561 M_1$, respectively; for $Q=0.3$, $l=0.05$, the two additional photon rings are located at $b_1 \simeq .9942 M_1$ and $b_1 \simeq 4.7456 M_1$, respectively. In radiation model $B$, for $Q=0.3$, $l=0.01$, the additional lensing ring is located between $b_1 \simeq 3.8221 M_1$ and $b_1 \simeq 5.056 M_1$; for $Q=0.3$, $l=0.05$, the additional lensing ring is located between $b_1 \simeq 3.9686 M_1$ and $b_1 \simeq 5.0764 M_1$. The results indicate that as the charge $Q$ and the Lorentz-violation parameter $l$ increase, the range of light bands outside the shadow decreases, but the range of specific additional rings increases. Therefore, the trend of change in the external shadow region is the same as that of the KR field BH, while the trend of change in specific additional rings is opposite.

\section{Conclusion and Discussion}\label{sec4}

This paper primarily investigates the observational characteristics of the ATSW and BH under the same mass parameter $M$, charge $Q$, and Lorentz-violation parameter $l$ in the KR field. Unlike BH, the ATSW connects two spacetimes with mass parameters $M_1$ and $M_2$ through a throat, assuming that the observer at infinity is located in the spacetime with mass parameter $M_1$. Initially, we calculate the variations in the photon sphere radius and critical impact parameter $b_c$ in the ATSW spacetime under different values of charge $Q$ and Lorentz-violation parameter $l$, as shown in Table.~(\ref{T1}) and Table.~(\ref{T2}). The results indicate that as the charge $Q$ and Lorentz-violation parameter $l$ increase, the photon sphere radius $r_p$ and critical impact parameter $b_c$ decrease accordingly. Next, we plot the relationship between the effective potential energy of the ATSW and BH and the variations in charge $Q$ and Lorentz-violation parameter $l$, as illustrated in Fig.3 and 4. The photon trajectories can be categorized into three scenarios: when $b_1<Zb_{c_2}$, photons fall from the $M_1$ spacetime into the $M_2$ spacetime and move to infinity; when $Zb_{c_2}<b_1<b_{c_1}$, photons enter the potential barrier of the $M_2$ spacetime from the $M_1$ spacetime and then return to the $M_1$ spacetime; when $b_1>b_{c_1}$, photons reach the potential barrier of the $M_1$ spacetime and then move towards infinity in the $M_1$ spacetime.

To better understand the photon trajectories and deflection angles, we have plotted the photon trajectories and deflection angles under different values of charge $Q$ and Lorentz-violation parameter $l$, as shown in Figs.5 and 6. The results indicate that $b_{c_1}$ decreases with the increase of $Q$ and $l$, whereas $Zb_{c_2}$ exhibits an opposite trend. It is also observed that the photon trajectories are significantly influenced by the Lorentz-violation parameter $l$. Additionally, we have calculated the transfer functions of ATSW and found that, unlike BH, ATSW possesses new second and third transfer functions, which correspond to the lens ring and photon ring group, respectively. The unique significance of the new transfer functions in the ATSW spacetime is that when $Zb_{c_2}<b_1<b_{c_1}$, the spacetime $M_2$ can return photons to the spacetime $M_1$.

Finally, we investigated the observational characteristics of ATSW and BH under two radiation models of a thin accretion disk. In radiation model $A$, unlike BH, ATSW exhibits two additional photon rings near the critical curves at $b_1 \simeq 3.7221 M_1$ and $b_1 \simeq 4.9676 M_1$. In radiation model $B$, unlike BH, ATSW presents an additional lens ring between the critical curves at $b_1 \simeq 3.7164 M_1$ and $b_1 \simeq 5.1139 M_1$. Therefore, these distinct observational features may provide an important basis for distinguishing ATSW from BH in observations.

\section*{Acknowledgments}
 This work is supported by the Sichuan Science and Technology Program (2024NSFSC1999).


\end{document}